\documentclass[journal=jacsat,manuscript=article]{achemso}

\usepackage[version=3]{mhchem} 
\usepackage{amsmath,amsfonts,amssymb}
\usepackage{braket}
\usepackage{xcolor} 
\usepackage{subcaption} 
\usepackage{titlesec} 
\usepackage{enumitem} 
\usepackage{hyperref}
\usepackage{array}

\hypersetup{colorlinks = True, allcolors = {blue}}

\newcommand{\revise}[1]{\textcolor{black}{#1}}

\newcommand{\BC}[1]{\hat{a}_{#1}^\dagger}
\newcommand{\BA}[1]{\hat{a}_{#1}}

\newcommand{\Eq}[1]{Eq.~({#1})}
\newcommand{\Fig}[1]{Figure~{#1}}

\newcommand{\Sec}[1]{Section~{#1}}

\author{Anthony M. Smaldone}
\affiliation{Department of Chemistry, Yale University, New Haven, CT 06520, USA}
\altaffiliation{These authors contribute equally to this article.}
\author{Yu Shee}
\affiliation{Department of Chemistry, Yale University, New Haven, CT 06520, USA}
\altaffiliation{These authors contribute equally to this article.}
\author{Gregory W. Kyro}
\affiliation{Department of Chemistry, Yale University, New Haven, CT 06520, USA}
\author{Chuzhi Xu} 
\affiliation{Department of Chemistry, Yale University, New Haven, CT 06520, USA}
\author{Nam P. Vu}
\affiliation{Department of Chemistry, Lafayette College, Easton, PA 18042, USA}
\alsoaffiliation{Department of Chemistry, Yale University, New Haven, CT 06520, USA}
\author{Rishab Dutta}
\affiliation{Department of Chemistry, Yale University, New Haven, CT 06520, USA}
\author{Marwa H. Farag}
\affiliation{Quantum Algorithm Engineering, NVIDIA Corporation, Santa Clara, CA 95051, USA}
\author{Alexey Galda}
\affiliation{Molecule Design and Modeling Group, Computational Science, Moderna Inc., 325 Binney Street, Cambridge, MA 02142}
\author{Sandeep Kumar}
\affiliation{Molecule Design and Modeling Group, Computational Science, Moderna Inc., 325 Binney Street, Cambridge, MA 02142}
\author{Elica Kyoseva}
\affiliation{Quantum Algorithm Engineering, NVIDIA Corporation, Santa Clara, CA 95051, USA}
\author{Victor S. Batista}
\email{victor.batista@yale.edu}
\affiliation{Department of Chemistry, Yale University, New Haven, CT 06520, USA}
\alsoaffiliation{Yale Quantum Institute, Yale University, New Haven, CT 06511, USA}

\title[An \textsf{achemso} demo]
  {\revise{Quantum Machine Learning in Drug Discovery: Applications in Academia and Pharmaceutical Industries}}

\abbreviations{QML}

\begin{document}







\begin{abstract}
    \revise{The nexus of quantum computing and machine learning––quantum machine learning––offers the potential for significant advancements in chemistry.} This review \revise{specifically} explores the potential of quantum neural networks on gate-based quantum computers \revise{within the context of drug discovery}. We \revise{discuss} the theoretical foundations of quantum machine learning, including data encoding, variational quantum circuits, and hybrid quantum-classical approaches. \revise{Applications to drug discovery are highlighted, including molecular property prediction and molecular generation.} We provide a balanced perspective, emphasizing both the potential benefits and the challenges that must be addressed.
\end{abstract}

\section{Introduction}

\subsection{Quantum Computing}
In this introduction, we discuss the general methodology of quantum computing based on unitary transformations (gates) of quantum registers, which underpin the potential advancements in computational power over classical systems. We introduce the unique properties of quantum bits, or qubits, quantum calculations implemented by algorithms that evolve qubit states through unitary transformations, followed by measurements that collapse the superposition states to produce specific outcomes, and lastly the challenges faced in practical quantum computing limited by noise, with hybrid approaches that integrate quantum and classical computing to address current limitations. This introductory discussion sets the stage for a deeper exploration into quantum computing for machine learning applications in subsequent sections.

Calculations with quantum computers generally require evolving the state of a quantum register by applying a sequence of pulses that implement unitary transformations according to a designed algorithm. A measurement of the resulting quantum state then collapses the coherent state, yielding a specific outcome of the calculation. To obtain reliable results, the process is typically repeated thousands of times, with averages taken over all of the measurements to account for quantum randomness and ensure statistical accuracy. This repetition is essential to achieve convergence, as each individual measurement only provides probabilistic information about the quantum state.

Quantum registers are commonly based on qubits. Like classical bits, qubits can be observed in either of two possible states (0 or 1). However, unlike classical bits, they can be prepared in superposition states, representing both $0$ and $1$ simultaneously with certain probability. In fact, the state of a single qubit can be described using the ket notation, as follows: 
\begin{equation}
    |\psi\rangle = \alpha|0\rangle + \beta|1\rangle,
    \label{eq:single_qubit_state}
\end{equation}  
where $\alpha$ and $\beta$ are complex amplitudes satisfying the normalization condition $|\alpha|^2 + |\beta|^2 = 1$. Such a state represent the states $|0\rangle$ and $|1\rangle$ simultaneously with probability $|\alpha|^2$ and \revise{$|\beta|^2$}, respectively. 

Quantum registers with $n$ qubits represent states that are linear combinations of tensor products of qubit states. Therefore, a register with $n$ qubits represents $2^n$ states simultaneously, offering a representation with exponential advantage over classical registers. For instance, the state of a register with two qubits represents four states simultaneously, as follows:
\begin{equation}
|\psi\rangle = \alpha_{00}|0 \rangle \otimes |0\rangle + \alpha_{01}|0\rangle \otimes |1\rangle + \alpha_{10}|1\rangle \otimes |0\rangle + \alpha_{11}|1\rangle \otimes |1\rangle,
\end{equation} 
with complex coefficients $\alpha_{jk}$ satisfying the normalization condition $|\alpha_{00}|^2 + |\alpha_{01}|^2 + |\alpha_{10}|^2 + |\alpha_{11}|^2 = 1$, and defining the probabilities $P_{jk} =|\alpha_{jk}|^2$ of observing the state collapsed onto state $|j \rangle \otimes |k \rangle$ when measuring the two qubits. 

Quantum gates, analogous to classical logic gates, are used to represent the effect of the pulses that manipulate the states according to unitary transformations. Commonly used gates for transformation of a single qubit are the gates represented by the Pauli matrices:
\begin{equation}
X = \begin{pmatrix}
0 & 1 \\
1 & 0
\end{pmatrix}, \quad
Y = \begin{pmatrix}
0 & -i \\
i & 0
\end{pmatrix}, \quad
Z = \begin{pmatrix}
1 & 0 \\
0 & -1
\end{pmatrix}.
\end{equation}
For example, the $X$ (or, NOT) gate, flips the state of a qubit from $|0\rangle$ to $|1\rangle$, and vice-versa. Another important class of transformations of a single qubit are the rotation gates $R_x(\theta)$, $R_y(\theta)$, and $R_z(\theta)$. The rotation around the $Y$-axis, for instance, is expressed as:
\begin{equation}
R_y(\theta) = \exp\left(-i \frac{\theta}{2} Y\right) = \cos\left(\frac{\theta}{2}\right)I - i\sin\left(\frac{\theta}{2}\right)Y,
\end{equation}
 where $I$ is the identity matrix. 
 
 For multi-qubit systems, universal computing can be achieved with single qubit gates (such as Pauli, or rotation gates) and the two-qubit CNOT (Controlled-NOT) gate, defined as follows:
\begin{equation}
\text{CNOT} = \begin{pmatrix}
1 & 0 & 0 & 0 \\
0 & 1 & 0 & 0 \\
0 & 0 & 0 & 1 \\
0 & 0 & 1 & 0
\end{pmatrix}.
\end{equation}

Measurements of individual qubits project the superposition states onto one of the basis states of the operator used for the measurement. Averages over many measurements (i.e., many shots) are required to achieve statistical converge of the calculation. For example, for a single qubit prepared in state $|\psi \rangle$, given by Eq.~\ref{eq:single_qubit_state}, measurements with the $Z$ operator yield either $1$ when the state is collapsed by the measurement onto state \( |0\rangle \) (with probability \( |\alpha|^2 \)), or $-1$ when the state is is collapsed into state $|1\rangle$ (with probability \( |\beta|^2 \)). 

\revise{Quantum mechanics introduces concepts such as superposition and entanglement, enabling computational parallelism. Quantum superposition allows for the representation and manipulation of an exponential number of states simultaneously, offering potential quantum advantage. Likewise, quantum entanglement, a form of correlation not present in classical systems, could further enhance this advantage.} The simplest example is a register prepared in the Bell state $|\psi \rangle = \alpha_{00} |0\rangle \otimes |0\rangle + \alpha_{11} |1\rangle \otimes |1\rangle$ where measurement of one of the two qubits collapses the state of {\em both} qubits in highly correlated way so that both qubits end-up in the same collapsed state (both $|0\rangle$, or both $|1\rangle$, but never one $|0\rangle$ and the other $|1\rangle$). These quantum properties offer great potential for computational advantage over classical computers and thus could lead to significant advancements in many areas of chemistry, and beyond.

Quantum algorithms can achieve significant speed up compared to their classical counterparts. For example, the Quantum Fourier Transform (QFT) \cite{coppersmith_approximate_2002} can enable \revise{exponential} speedup when compared to the best-known classical Fourier  transform algorithms. Algorithms like Quantum Phase Estimation \cite{kitaev_quantum_1995} and the Shor's algorithm\cite{shor_algorithms_1994,shor_polynomial-time_1997} can also enable factorization of large numbers with exponential quantum advantage. Amplitude amplification techniques, such as those used in Grover's algorithm, provide a quadratic speed up for unstructured search problems, while the Harrow-Hassidim-Lloyd (HHL) algorithm \cite{harrow_quantum_2009} offers logarithmic speed up for solving linear systems within bounded error, highlighting the potential of quantum computing to outperform classical methods in a wide range of applications. The actual implementation of these quantum algorithms, however, would require fault-tolerant quantum computers that are not currently available to achieve quantum advantage over classical algorithms.

Due to the current limitations of quantum hardware, including noise and limited qubit counts, significant efforts have been focused on near-term calculations based on hybrid quantum-classical approaches where only part of the calculation is performed on the quantum computer while the rest of the computation is delegated to conventional high-performance computers. For example, variational algorithms, such as the Variational Quantum Eigensolver (VQE) \cite{peruzzo_variational_2014} and Quantum Imaginary Time Evolution (QITE),~\cite{moll2018quantum,mcardle2019variational,yuan2019theory,gomes2020efficient,Motta2020determining,nishi2021implementation,selvarajan2021prime,huang2022efficient,yeter2022quantum,kyaw2023boosting} 
implement hybrid quantum-classical approaches. These algorithms generate quantum states and employ classical computations to combine the results of the measurements performed on the quantum states. This synergy leverages the strengths of both quantum and classical resources, making it feasible to solve problems with current noisy intermediate-scale quantum (NISQ) devices.

Despite significant advances in the field, an outstanding challenge is to achieve advantage over classical high performance computing. One promising direction is the use of quantum computers to implement machine learning algorithms. Harnessing the speed-up of quantum algorithms could address complex problems in data analysis and pattern recognition. \revise{Given that quantum computing has many potential applications in chemistry and biological science, there is a great deal of hope that quantum machine learning (QML) can be extended to these areas of research.}

\subsection{Machine Learning}

Machine learning \revise{(ML)} algorithms are able to {\em learn} from data, where \textit{learning} in this context can be defined according to ``a program is said to learn from experience \textit{E} with respect to some other class of tasks \textit{T} and performance measure \textit{P}, if its performance at tasks in \textit{T}, as measured by \textit{P}, improves with experience \textit{E}" \cite{Goodfellow-et-al-2016}. In practice, machine learning can be used to approximate a function of the input data to predict some variable (e.g., predict chemical toxicity from molecular features)~\cite{smaldone_quantum--classical_2024, kyro_cardiogenai_2024}, or can be used to learn the distribution of the input data to generate synthetic data akin to the training distribution (e.g., generating virtual compounds with specific drug-like properties) \cite{kyro_chemspaceal_2024}.

There are now many machine learning methods that have demonstrated exceptional, unprecedented abilities in many areas of research pertaining to drug development, with AlphaFold 2 and its later iterations being particularly recognizable  \cite{jumper_highly_2021,abramson_accurate_2024}. AlphaFold is able to predict protein structures from their input sequences with high accuracy, although it is less capable in cases where the input sequence corresponds to a structure that is not well represented in the training distribution. Nonetheless, there is a lot of excitement and anticipation that AlphaFold will enable a lot of innovation within the domains of studying protein dynamics and hit identification in drug discovery \cite{varadi_impact_2023}.

Machine learning has become pervasive in cheminformatics, and there have been many tools developed to predict molecular properties, generate compounds with prespecified properties, and ultimately reduce an incredible vast chemical search space to something tractable given the specific task at hand \cite{niazi_recent_2023}. Specifically, there are a lot of efforts leveraging machine learning to reveal molecular mechanisms \cite{paul_mldspp_2024}, analyze complex biochemical data \cite{zhou_deep-cloud_2024}, process and optimize chemical data \cite{khashei_intelligent_2023}, predict protein structure \cite{jumper_highly_2021,abramson_accurate_2024}, virtual screening and drug design \cite{kyro_hac-net_2023, oliveira_virtual_2023}, protein-ligand docking \cite{corso_diffdock_2023}, as well as many other tasks \cite{schrier_pursuit_2023}.

\subsection{Quantum Neural Networks} \label{QNN_intro}

\revise{Machine learning has had a transformative effect on all facets of modern life and has led to increasing computational demands, thus motivating the development of QML methods.} The promising capabilities of quantum computing have already motivated the development of quantum analogs for a wide range of classical machine learning methods. Bayesian inference\cite{low_quantum_2014}, least-squares fitting\cite{wiebe_quantum_2012}, principal component analysis \cite{lloyd_quantum_2014}, and support vector machines \cite{rebentrost_quantum_2014} are some of the algorithms for which quantum counterparts have already been developed. While quantum analogs for these traditional ML methods have a demonstrable quantum speed-up,~\cite{biamonte_quantum_2017} some of the most awe-inspiring advances are due to artificial neural networks (ANNs). 

Perhaps the earliest discussions of quantum neural networks (QNNs) were motivated by studies of neural function through the lens of quantum mechanics \cite{kak_quantum_1995}. Since then, the field has evolved to exploiting the computational parallelization enabled by superposition states and entanglement \cite{panella_neural_2011}. In the early stages of research for QNNs, much effort was dedicated towards developing quantum systems that preserved the mechanisms of classical ANNs \cite{gupta_quantum_2001,zhou_quantum_2012,goncalves_quantum_2014}. However, those efforts have largely failed to reconcile the linear dynamics of a quantum state evolving through a circuit and the non-linear behavior of classical neural networks \cite{schuld_quest_2014}. Increasingly, the field has consolidated around the use of variational quantum circuits to learn data representations \cite{farhi_classification_2018} rather than directly creating a quantum analog of a neural network. Accordingly, quantum versions of the most popular classical neural network architectures, such as convolutional neural networks (QCNNs), graph neural networks (QGNNs), variational autoencoders (QVAEs), and generative adversarial networks (QGANs) have been realized and centered around variational quantum circuits.

\begin{figure}[!h]
    \centering
    \includegraphics[width=0.5\textwidth]{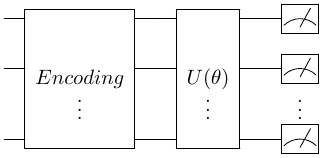}
    \caption{Structure of a typical quantum neural network. The input is encoded into a quantum state, followed by a variational quantum circuit and measurements.}
    \label{fig:quantum_neural_network_structure}
\end{figure}

\revise{QNN}s require data encoding, variational quantum gates with learnable parameters $\theta$, and measurements, as depicted in Figure \ref{fig:quantum_neural_network_structure}. Data encoding converts classical data into a quantum state. The choice of strategy for data encoding can be of paramount importance in QNNs, as it can \revise{significantly} affect performance and impact the underlying computational complexity. While other data encoding strategies exist\cite{larose_robust_2020, pande_comprehensive_2024}, three of the most popular methods are discussed in the following subsections.

\subsubsection{Basis Encoding}
\begin{figure}[htbp]
  \centering
  \begin{subfigure}[b]{0.30\textwidth}
    \centering
    \includegraphics[height=3cm]{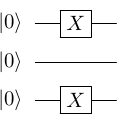}
    \caption{basis encoding}
    \label{fig:basis_encoding_circuit}
  \end{subfigure}
  \hfill
  \begin{subfigure}[b]{0.30\textwidth}
    \centering
    \includegraphics[height=3cm]{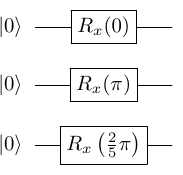}
    \caption{angle encoding}
    \label{fig:angle_encoding_circuit}
  \end{subfigure}
  \hfill
  \begin{subfigure}[b]{0.30\textwidth}
    \centering
    \includegraphics[height=3cm]{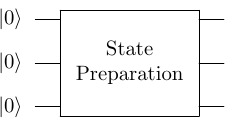}
    \caption{amplitude encoding}
    \label{fig:amplitude_encoding_circuit}
  \end{subfigure}
  \caption{Data encoding methods. (a) Quantum circuit to prepare the [1,0,1] vector with basis encoding. (b) Quantum circuit to prepare the pre-scaled $[0, \pi, \frac{2\pi}{5}]$ vector with angle encoding, choosing the x-rotation axis. (c) Amplitude encoding is equivalent to quantum state preparation.}
  \label{fig:}
\end{figure}

Basis encoding is a straightforward and inexpensive method to encode binary data into a quantum system. Explicitly, let $\mathcal{D}$ be a classical \revise{binary} dataset such that each element $x^m \in \mathcal{D}$ \revise{is} an $N$-bit binary string of the form $x^m = (b^m_1, b^m_2, \cdots, b^m_N)$, with $b^m_j= 0$ or $1$. Then the classical dataset can be represented by the quantum state $\ket{\mathcal{D}}$ of $N$ qubits, \revise{where $M$ is the total number of basis states used for the encoding}:
\begin{equation}
    \ket{\mathcal{D}} = \frac{1}{\sqrt{M}}\sum_{m=1}^M\ket{x^m}.
    \label{basis_encoding}
\end{equation}
where $x^m$ corresponds to the $m$-th element of the data set. For encoding a specific element (e.g., the binary string $[1,0,1]$) we simply place a Pauli $X$ gate on the qubits that should be one (e.g., on the first and third qubits, as shown in Figure \ref{fig:basis_encoding_circuit}).

\subsubsection{Angle Encoding}\label{angle_encoding}
Unlike basis encoding where the data is restricted to binary values, angle encoding allows data to take the form of real, floating point numbers. This encoding method entails rotating the state of a qubit around an axis of the Bloch sphere by an angle corresponding to the classical data. Explicitly, for an element $x^m$ of a classical dataset $\mathcal{D}$ where $x^m \in [0,2\pi]$, then the value of $x^m$ may be encoded into a single qubit by a rotation operator:

\begin{equation}
    x \rightarrow R_k(x)\ket{0} = e^{-ix\sigma_k/2}\ket{0},
    \label{angle_encoding}
\end{equation}
where $k$ indicates the rotation axis ({\em e.g.}, $k = y$).

Classical datasets seldom satisfy the $2\pi$-periodicity requirement of rotation gates. Nevertheless, the data can always be normalized such that $x^m \in [0,2\pi]$, or commonly $x^m \in [0,\pi]$. For example, suppose the task is to encode the vector $x = [0,5,2]$ into a quantum state via angle encoding with a maximum rotation angle of $\pi$. The vector is first normalized to the range [0,$\pi$]:
\begin{equation}
    x_{angles} = \pi \cdot \frac{x-\min(x)}{\max{(x)}-\min(x)} = [0, \pi, \frac{2\pi}{5}].
    \label{angle_normalization}
\end{equation}
After scaling, the angles can be encoded with the $R_x$ or $R_y$ gates, as shown in Figure \ref{fig:angle_encoding_circuit}.


\subsubsection{Amplitude Encoding} \label{sec: amp_encoding}
Amplitude encoding allows one to encode complex valued floats into the amplitudes of a quantum state. Thus, for a given classical dataset $\mathcal{D}$, an L2-normalized complex vector $x \in \mathcal{D}$ of length $N$ can be encoded into $\log(N)$ qubits. Namely,

\begin{equation}
    x \rightarrow U_x\ket{0}^{\otimes _{\log(N)} } = \ket{x} = \sum_{k=0}^{2^N-1}\alpha_k\ket{k}.
    \label{amplitude_encoding}
\end{equation}

Many quantum neural networks rely on this encoding strategy, as it enables an exponential reduction in the number of required bits to represent data, and thus has the potential to allow for a speed-up that is not possible on classical computers. Despite this, the unitary operator $U_x$ shown in Equation \ref{amplitude_encoding} may demand a significant number of gates - a challenge discussed further in section \ref{challenges_algorithms}.

\subsubsection{Variational Quantum Circuits and Readout} \label{sec: variational_quantum_circuits_readout}

\begin{figure}[!h]
    \centering
    \includegraphics[width=0.6\textwidth]{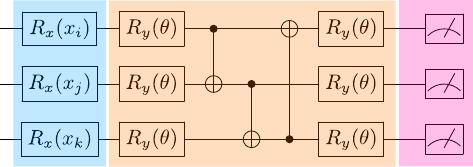}
    \caption{Generic three qubit quantum neural network using x-axis angle encoding (blue), variational quantum circuit (orange), and measurements (pink).}
    \label{fig:quantum_neural_network_colored}
\end{figure}

\begin{figure}[!h]
    \centering
    \includegraphics[width=0.6\textwidth]{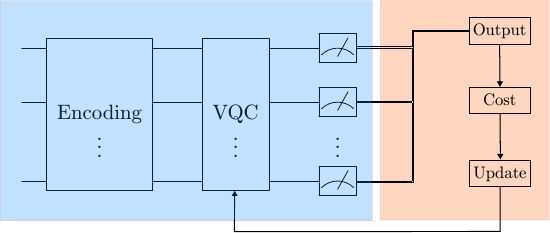}
    \caption{Training a general variational quantum circuit (blue) via classical post-processing and optimization (orange).}
    \label{fig:training_VQC}
\end{figure}

 Variational quantum circuits (VQCs), also commonly known as parameterized quantum circuits (PQCs), are typically used to introduce learnable parameters $\theta$ of unitary gates (Fig.~\ref{fig:quantum_neural_network_colored}).
 After the VQC, measurements are performed. Measurements typically undergo classical post-processing to obtain averages. The \revise{set of} parameters \( \theta \) are iteratively adjusted by a classical computer to minimize a cost function $C(\theta)$ defined by the average expectation values $\langle \phi | U^\dagger(\theta) \hat{O} U(\theta) | \phi \rangle$, as follows: 
\begin{equation}
    C(\theta) = f\left(\langle \phi | U^\dagger(\theta) \hat{O} U(\theta) | \phi \rangle\right),
    \label{cost_function}
\end{equation}
where $| \phi \rangle$ are the encoded states and $U(\theta)$ is the ansatz of choice with learnable parameters \revise{and the} function $f$ \revise{is any classical post-processing function}.
The overall hybrid quantum-classical machine learning scheme is depicted in Figure \ref{fig:training_VQC}.

\section{Predictive Quantum Machine Learning}
\subsection{Quantum Graph Neural Networks}

\begin{figure}[!h]
    \centering
    \includegraphics[width=1.0\textwidth]{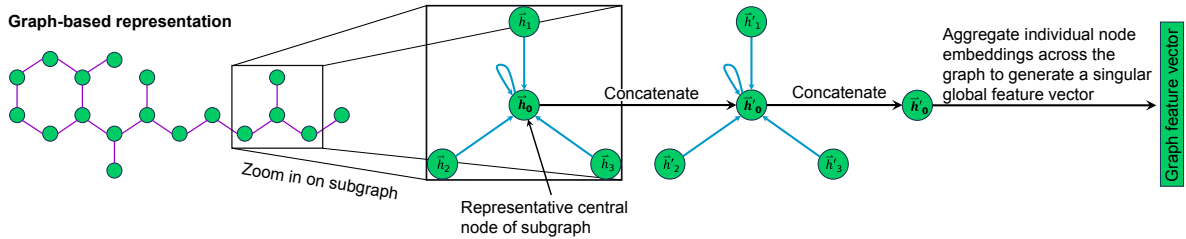}
    \caption{A classical graph neural network for extracting features from a molecule.}
    \label{fig:classical_gnn}
\end{figure}

Graph neural networks (GNNs) are popular models in applications of machine learning methods to chemistry because molecules can be intuitively represented as graphs where nodes are atoms and edges are bonds (Figure~\ref{fig:classical_gnn}). In a typical GNN, messages ({\em i.e.}, features used to describe each node) are passed between neighboring nodes, ultimately resulting in an aggregated graph-level encoding which can subsequently be processed to predict some value (e.g., protein-ligand binding affinity, hERG activity, etc.)~\cite{reiser_graph_2022}

QGNNs were first introduced with the Networked Quantum System.~\cite{verdon_quantum_2019}
In this system, a graph $\mathcal{G} = \{\mathcal{V}, \mathcal{E}\}$ with the set of nodes $\mathcal{V}$ and edges $\mathcal{E}$ is defined as tensor products of Hilbert subspaces representing nodes and edges. The Hilbert space representing nodes, $\mathcal{H}_\mathcal{V}= \bigotimes_{v \in \mathcal{V}}\mathcal{H}_v$ and the space representing edges $\mathcal{H}_\mathcal{E}= \bigotimes_{e \in \mathcal{E}}\mathcal{H}_e$ are joined to create the full networked Hilbert space $H_\mathcal{G} =\mathcal{H}_\mathcal{V} \otimes \mathcal{H}_\mathcal{E}$ that comprise the space for the complete graph. Since then, various quantum theoretical formulations of QGNNs have been introduced.~\cite{zheng_quantum_2021, tuysuz_hybrid_2021, beer_quantum_2021, liao_graph_2024} Alongside quantum graph convolutional networks, quantum learning on equivariant graphs has also been demonstrated \cite{mernyei_equivariant_2022,skolik_equivariant_2023}, which has been of increasing interest in classical ML for drug discovery \cite{chen_knowledge_2024,cremer_equivariant_2023,dhakal_predicting_2023}. 

Equivariant QGNNs and hybrid quantum-classical QGNNs have been used to predict the HOMO-LUMO gap in the QM9 dataset, which can provide insights on molecular stability.~\cite{ryu_quantum_2023} An interesting observation \revise{from this work} is that comparisons of \revise{their} QGNN models to their corresponding classical models with the same number of parameters shows that the quantum models typically outperform the classical counterparts. Additionally, training of the quantum model is generally more efficient. These are exciting results that suggest favorable scalability and generalization of QGNNs, as previously suggested.~\cite{caro_generalization_2022} Another study,\cite{vitz_hybrid_2024} has implemented a hybrid QGNN to predict the formation energy of perovskite materials. While their method underperforms compared to the fully classical GNN, it has been pointed out that advantages will emerge once state preparation techniques improve due to their usage of amplitude encoding. 

Quantum isomorphic graph networks and quantum graph convolutional networks have been used to predict protein ligand binding affinities, showing that hybrid models already perform on par with state-of-the-art models.~\cite{dong_prediction_2023} In this work, features are amplitude encoded into a quantum state and a PQC replaces the classical \revise{multi-layer perceptron (MLP)} to perform convolutions. The models provide a good balance between number of parameters and generalization.

QGNNs are truly promising methods. For example, Liao \textit{et al.} \cite{liao_graph_2024} has analyzed quantum implementations of the Simple Graph Convolutional network \cite{wu_simplifying_2019} and the linear graph convolutional network \cite{pasa_empowering_2024} that exhibit quantum advantage in terms of both space and time complexity. As the utility of graph networks is extended to both small molecules and large protein structures alike, solutions with complexity advantages are expected to be the dominant driver of the success of QGNNs.

\subsection{Quantum Convolutional Neural Networks}
Convolutional neural networks (CNNs) gained initial popularity for their success in image detection and classification\cite{dhillon_convolutional_2020}. They have been applied in chemistry to predict molecular properties, interaction strengths, and other chemically significant tasks \cite{jiang_convolutional_2021}. The most fundamental architectural components of CNNs are the kernels of convolutional layers \cite{oshea_introduction_2015}. Each kernel creates a linear combination of the values in the spatial neighborhood of a given voxel (i.e., a pixel in the 2D case or a point in a 3D grid) of the input data and then propagates the resulting scalar to a corresponding spatial index in the output array. The coefficients for this linear combination are learned throughout training and constitute the weights of the kernel, which are applied uniformly across the input voxels.

QCNNs were first introduced for quantum phase recognition \cite{cong_quantum_2019}, outperforming existing approaches with a significantly reduced number of variational parameters, scaling as $O(\log(N))$ with $N$ the number of qubits. This initial success sparked significant interest, leading to the development of many QCNN variants \cite{kerenidis_quantum_2019, henderson_quanvolutional_2020, liu_hybrid_2021, maccormack_branching_2022, smaldone_quantum_2023}, tutorials \cite{oh_tutorial_2020, pennylane-qcnn, qiskit-qcnn,tf-qcnn}, and applications to a large range of complex tasks in many fields of science and technology. For example, in high energy physics, QCNNs have been used to classify particles with a level of accuracy and speed of convergence that surpasses classical Convolutional Neural Networks (CNNs) with the same number of learnable parameters \cite{chen_quantum_2022}. In the field of biochemistry, they have shown the ability to generate protein distance matrices \cite{hong_quantum_2021} and predict protein-ligand binding affinities \cite{domingo_binding_2023, dong_prediction_2023}, demonstrating their potential to contribute to our understanding of complex biological systems. 

\begin{figure}[!h]
    \centering
    \includegraphics[width=0.8\textwidth]{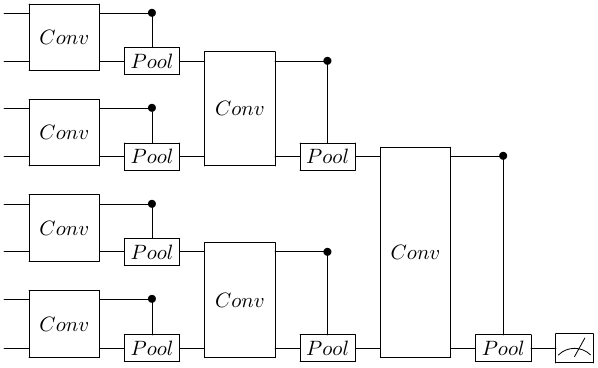}
    \caption{QCNN architecture introduced by Cong \textit{et al.} \cite{cong_quantum_2019} with $\log(N)$ parameters, where $N$ is the number of qubits.}
    \label{fig:QCNN_circuit}
\end{figure}

The appeal of QCNNs over many quantum counterparts of classical neural networks is multi-faceted. In a QCNN, the classical convolutional filters are replaced by quantum circuits (Figure~\ref{fig:QCNN_circuit}). 
In CNNs, the computation involves the discrete convolution between a relatively small kernel and the input data. This is attractive,  as it allows the quantum approach to load only a small amount of information at a time onto quantum devices, as determined by the kernel size, which is of paramount importance during the NISQ era. This feature of QCNNs can be particularly useful in a biological context, as full-size feature maps would be too demanding. 

Broadly speaking, there are two classes of QCNNs that could offer quantum advantage. This first class is akin to the general structure shown in Figure \ref{fig:QCNN_circuit}.~\cite{cong_quantum_2019}
QCNNs with that structure incorporate pooling layers that halve the number of active qubits with each successive layer. This architectural choice involves only $O(\log(N))$ parameters and effectively circumvents the issue of barren plateaus —a significant challenge discussed further in Section \ref{challenges_algorithms}. The second class can be termed Hybrid-QCNNs (HQCNNs). HQCNN models replace the forward pass of a convolutional filter with a quantum circuit, but perform pooling layers classically after a measurement. HQCNNs are popular choices since they allow for more classical control over the network, with the mixing of quantum and classical components potentially offering performance gains at the expense of trainability and complexity brought by the original QCNN architecture.

QCNNs and HQCNNs offer distinct advantages that are attractive for chemical and pharmaceutical applications. While QCNNs require only $O(\log(N))$ parameters and avoid barren plateaus, this by itself does not deem them to be advantageous over classical CNNs. In a rigorous analysis of QCNNs (to which they later extend to all QML models), the generalization bounds of these models were investigated \cite{caro_generalization_2022}. The reported analysis offers a guide to determine whether a QML model can exhibit better performance on unseen (test) data when compared to their classical counterpart. It is shown that when a QML model achieves a small training error on a given task, while the classical model with the same training error is significantly more complex, then the QML model will most likely outperform the classical model on unseen data. 

This simple guide is particularly useful for drug discovery applications where datasets can often be limited but good generalization is paramount for discovery of life-saving compounds. Given that in the NISQ era QML models can only include a limited number of parameters, it is commonplace and intuitive when designing QML models to compare their performance to a classical network of equal parameters. Therefore, it is important to temper claims of advantage in the event of comparing a quantum and classical models, wherein the classical model might be heavily restricted for the sole purpose of fair comparisons with equal number of parameters. Instead, it is more significant to identify tasks which satisfy the criteria which guarantee good generalization bounds\cite{caro_generalization_2022}. Shifting focus to this task identification, we anticipate that applications that are more likely to benefit from demonstrable quantum advantage are those for which the training data is scarce.

\begin{figure}[!h]
    \centering
    \includegraphics[width=1.0\textwidth]{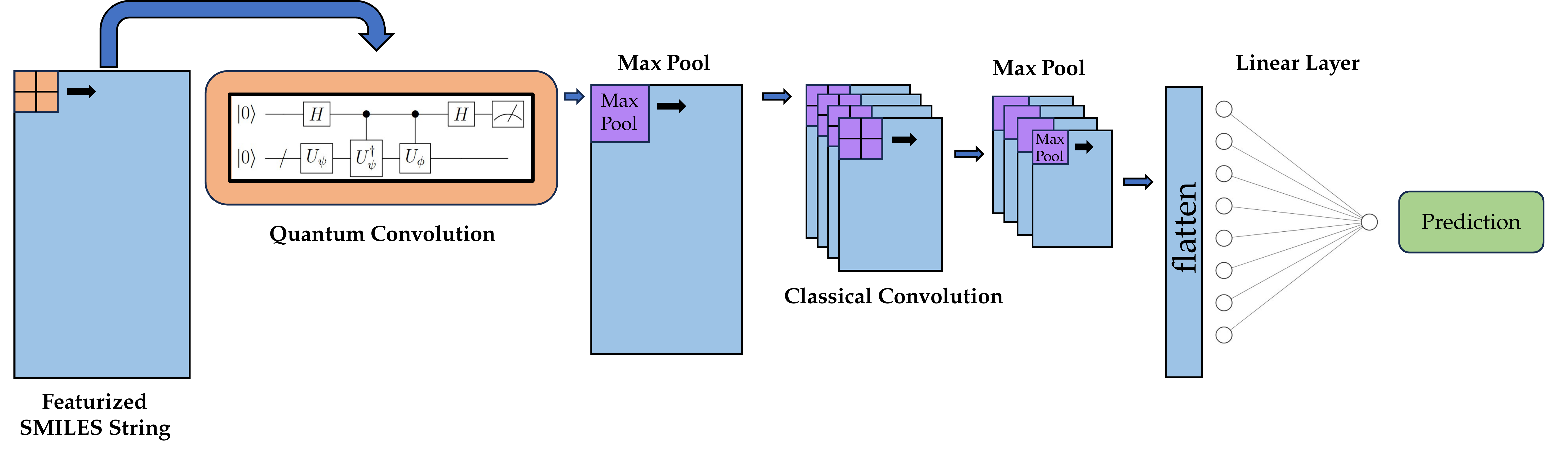}
    \caption{Hybrid Quantum-Convolutional Neural Network (HQCNN), adapted from Smaldone and Batista \cite{smaldone_quantum--classical_2024}. The full architecture contains both a quantum VQC layer, followed by classical pooling and classical convolutional layers. }
    \label{fig:hybrid_qcnn}
\end{figure}

HQCNNs operate differently and more flexibly than QCNNs. So, the ways in which quantum advantage might be demonstrated is likely different from QCNNs. While the above criteria to identify potential generalization quantum advantage would still apply to HQCNNs, this becomes less straightforward as HQCNNs do not necessarily operate with $O(\log(N))$ parameters like their fully quantum counterparts. HQCNNs have been proposed to enable quantum speed-up in the CNN architecture (and neural network architectures in general) by directly calculating the inner product of the filter and input data (Fig.~\ref{fig:hybrid_qcnn}) \cite{smaldone_quantum--classical_2024,stein_qucnn_2022,zhao_building_2019}. 
These approaches are attractive when searching for quantum advantage, as they are task-agnostic and the potential for realization on quantum hardware is dictated almost exclusively by data representation - a much more straightforward litmus test of advantage compared to that required for a generalizability advantage.

The success of classical CNNs in drug discovery has prompted the exploration of QCNNs, as in the domain of biophysics where the relatively large input data can be broken up into tractable quantum circuits using the HQCNN methodology. An early biophysical application of HQCNNs has involved a model capable of predicting protein structure,~\cite{hong_quantum_2021}
where the sequence lengths of the protein chains range from 50 to 500 residues and 50 to 266 residues in the training and testing sets, respectively. The reported results indicate commensurate performance to predictions by the popular classical model DeepCov \cite{jones_high_2018} for protein contact maps while offering faster training convergence. Both Domingo \textit{et al.} \cite{domingo_binding_2023} and Dong \textit{et al.} \cite{dong_prediction_2023} trained HQCNNs to predict protein-ligand binding affinities. Domingo \textit{et al.} demonstrated that their HQCNN architecture is able to reduce the number of parameters by 20\% while maintaining performance. They noted that depending on the hardware, this translates to a 20\% to 40\% reduction in training times. Similarly, Dong \textit{et al.} demonstrated competitive results with force ﬁeld-based MM/GBSA and MM/PBSA calculations while reducing the overall number of parameters to their classical counterparts. 

\begin{figure}[htbp]
  \centering
  \includegraphics[width=0.49\textwidth]{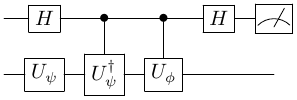}\\
  \includegraphics[width=0.49\textwidth]{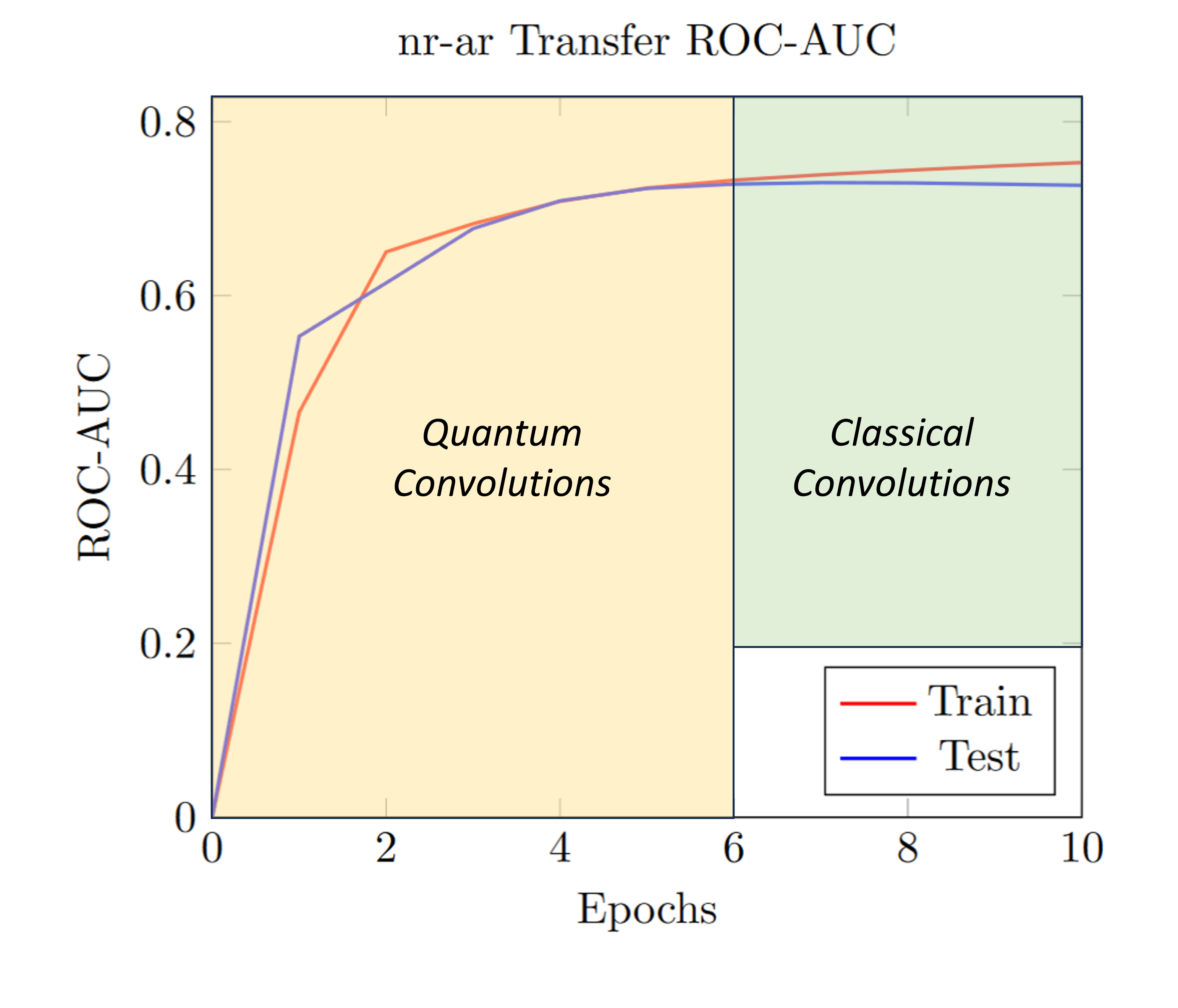}
  \caption{Quantum CNN Summary: (Top) Quantum circuit by Smaldone and Batista\cite{smaldone_quantum--classical_2024} employed to train a QCNN that predicts drug toxicities with a quadratic quantum speed-up for matrix multiplication. (Bottom) Learning curve for prediction of drug activity to the androgen receptor. The yellow region indicates epochs where the model was trained with reduced complexity using quantum circuits, and the green region shows where the weights derived and training was continued using a classical CNN.}
  \label{fig:quantum_cnn_summary}
\end{figure}

In the work by Smaldone and Batista \cite{smaldone_quantum--classical_2024}, a HQCNN has been trained to predict drug toxicity (Figure~\ref{fig:quantum_cnn_summary}). 
This work has demonstrated a method where the weights of a convolutional layer are learned via quantum circuits while performing the underlying matrix multiplication of discrete dot products with quadratic quantum speed-up. This strategy performs at the level of classical models with equal number of parameters and can be transferred to a classical CNN mid-training to allow for noiseless training convergence.

\subsection{Looking Ahead: Quantum Machine Learning for Large Molecules}

\begin{figure}[htbp]
    \centering
    \includegraphics[width=0.4\textwidth]{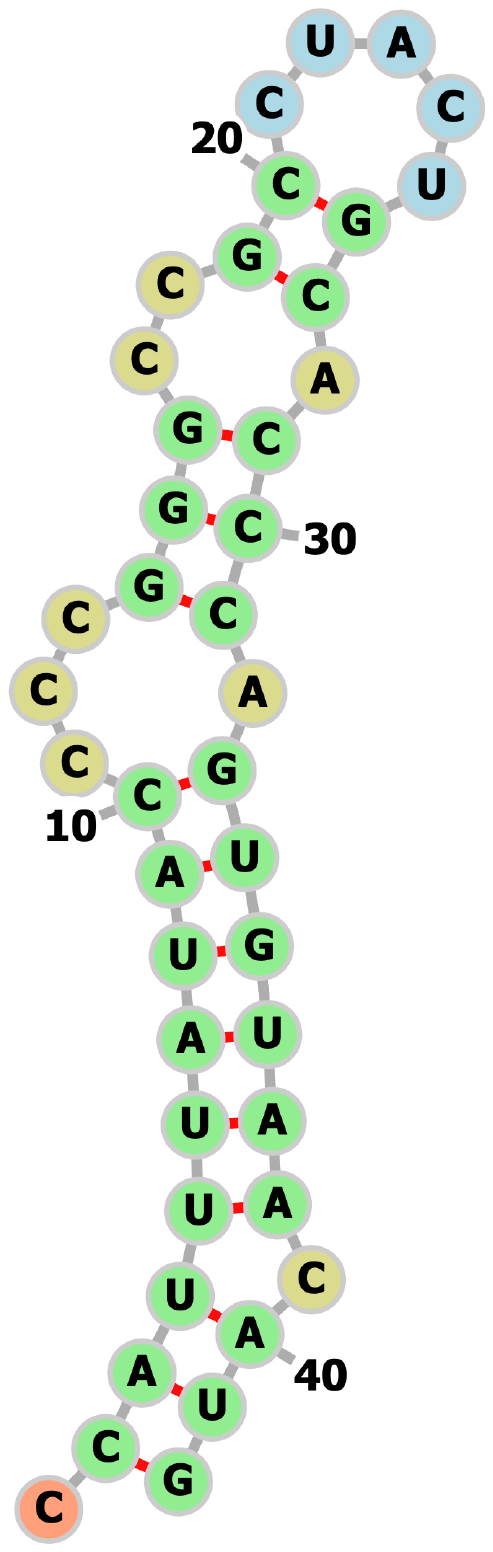}
    \caption{Optimal folded mRNA structure of the 42-nucleotide sequence computed using the VQE algorithm with 80 physical qubits on the IBM quantum processor Heron.}
    \label{fig:mrna_secondary}
\end{figure}

\revise{mRNA and antibody-based biotherapeutics are critical for the development of next-generation therapies, yet both pose complex challenges, such as determining mRNA structures and understanding antibody-antigen interactions. Quantum computing has already shown promise by predicting mRNA secondary structures  (see Figure \ref{fig:mrna_secondary}) \cite{alevras2024mrna}, and quantum neural networks are now being applied to tackle antibody-antigen interactions. Notably, Paquet et al. \cite{paquet_quantumbound_2024} introduced QuantumBound, a hybrid quantum neural network designed to predict the physicochemical properties of ligands within receptor-ligand complexes. Furthermore, Jin et al. \cite{jin_quantum_2022} developed a QNN model to predict potential COVID-19 variant strains using available SARS-CoV-2 RNA sequences. These early successes highlight the potential of quantum neural networks to address key challenges in biotherapeutics.}



\section{Generative Quantum Machine Learning}

\subsection{Quantum Autoencoders}

\begin{figure}[htbp]
  \centering
  \begin{subfigure}[b]{0.28\textwidth}
    \centering
    \includegraphics[width=\textwidth]{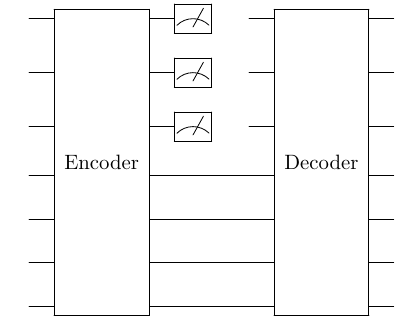}
    \caption{QAE.}
    \label{fig:qae_architecture}
  \end{subfigure}
  \hfill
  \begin{subfigure}[b]{0.55\textwidth}
    \centering
    \includegraphics[width=\textwidth]{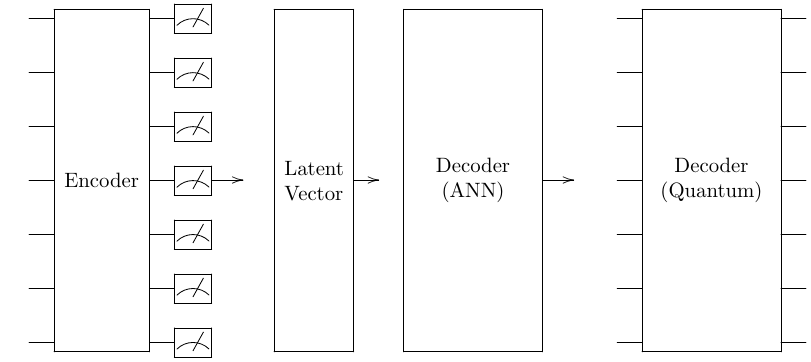}
    \caption{HQA.}
    \label{fig:hqa_architecture}
  \end{subfigure}
  \hfill
  \begin{subfigure}[b]{0.55\textwidth}
    \centering
    \includegraphics[width=\textwidth]{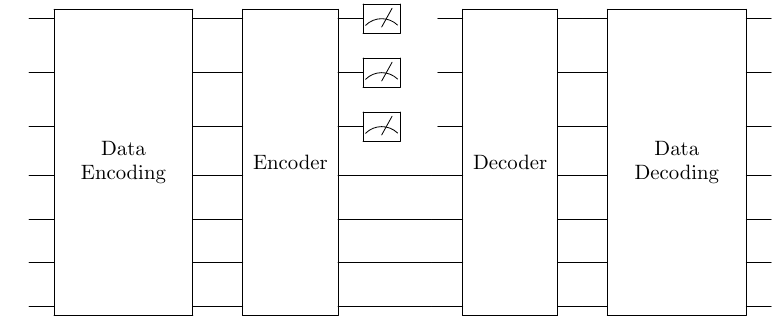}
    \caption{QAE for classical data.}
    \label{fig:qae_classical_data_architecture}
  \end{subfigure}
  \caption{Different types of Quantum Autoencoder (QAE). (a) QAE utilizing a fully quantum circuit as the model architecture. (b) Hybrid Quantum Autoencoder (HQA) with classical latent representation. (c) QAE with classical data as input and output.}
  \label{fig:qae_summary}
\end{figure}

The primary motivation behind the development of autoencoders is to compress data into a latent space, reducing dimensionality while preserving essential information of the training data. Similarly, the original motivation for development of quantum autoencoders (QAEs) is to compress quantum data (Figure~\ref{fig:qae_summary}). 
Variational Autoencoders (VAEs)\revise{, a specific type of autoencoders,} have gained popularity for molecular generation due to their ability to learn compact representations of molecular structures and generate new molecules with similar properties. \revise{QVAEs can compress} quantum states \revise{and} could therefore \revise{enable} new avenues for molecular generation, though the exact benefits of QVAEs in this domain require further investigation.

There are two primary types of QAE, both utilizing hybrid quantum-classical schemes where classical computers are used for parameter optimization. The first type employs a quantum circuit as the model architecture~\cite{romero_quantum_2017, lamata_quantum_2018, bondarenko_quantum_2020, cao_noise-assisted_2021, pepper_experimental_2019, ding_experimental_2019, huang_realization_2020}, aiming to leverage quantum gates and operations to encode and decode quantum states (see Fig.~\ref{fig:qae_architecture}). The second type, known as the Hybrid Quantum Autoencoder (HQA)~\cite{srikumar_clustering_2022}, employs measurement outcomes as the latent representations. This approach combines classical networks with \revise{QNNs} in a hybrid model architecture, where classical vectors derived from quantum measurements are accessible for further analysis and processing (see Fig.~\ref{fig:hqa_architecture}). Note that the compression is effective (with no loss of information) only if the set of states to be compressed has support on a subspace (lower dimension) of its Hilbert space~\cite{ma_compression_2022}. For example, the success of the Hubbard model example from Romero et al~\cite{romero_quantum_2017}. is due to the fact that these physical states exhibit certain symmetries.

A proposal for QVAE~\cite{wang2024zeta} involves the model architecture of the first type of QAE shown in Fig.\ref{fig:qae_architecture} and a latent representation regularized as in classical VAE. The regularized latent space can enhance classification performance compared to QAE. However, the regularization process requires mid-circuit quantum state tomography, which may represent a practical challenge for fully characterizing the state and scaling up.

Despite the promising aspects of QAE, several challenges remain. First, training relies on classical optimization algorithms, which can obscure statements about the overall computational complexity. Second, these models assume that input states can be efficiently prepared, a relatively straightforward task for quantum data but challenging for classical data (see Fig.~\ref{fig:qae_classical_data_architecture}). The encoding of classical data into quantum states might negate the computational benefits offered by quantum computers. Consequently, no immediate advantage can be claimed for QAE on classical data over classical methods at present. However, advancements in quantum computing hardware and more efficient optimization schemes could lead to significant improvements, making QAE a more viable and efficient tool in the future, particularly as the training optimization and data encoding complexity becomes comparable to the quantum components of the models.

\subsection{Quantum Generative Adversarial Networks}

Generative Adversarial Networks (GANs) are machine learning models designed to generate new data samples that mimic samples from a given distribution. GANs consist of three primary components: the prior distribution/noise sampling, the generator, and the discriminator. The generator creates data samples from random noise \revise{sampling}, while the discriminator evaluates the authenticity of the generated samples by comparison against real data. This adversarial training process helps the generator improve over time, creating increasingly realistic samples. GANs have found applications in molecular generation, much like \revise{VAEs}, and have been shown to generate novel molecular structures that adhere to desired properties~\cite{guimaraes2017objective, sanchez2017optimizing, putin2018reinforced, putin2018adversarial}. 

A \revise{QGAN} was proposed by Dallaire-Demers and Killoran~\cite{dallaire-demers_quantum_2018}. This work introduced the concept of using quantum circuits within the GAN framework, specifically leveraging quantum circuits to measure gradients. Romero and Aspuru-Guzik~\cite{romero_variational_2021} extended the concept of QGANs by modeling continuous classical probability distributions using a hybrid quantum–classical approach. While their results were promising, they noted that further theoretical investigations were necessary to determine whether their methodology offers practical advantages over classical approaches.

QGANs have been applied to generation of small molecules,~\cite{li_drug_2021} in a study that applied QGANs to the QM9 dataset~\cite{ramakrishnan_quantum_2014}. That study reported better learning behavior due to the claimed superior expressive power and fewer parameters required by the quantum models. However, these QGANs struggled to generate valid molecules, and subsequent tests by other researchers indicated that these QGANs struggled to generate train-like molecules~\cite{kao_exploring_2023}.

Kao et al.~\cite{kao_exploring_2023} explored the advantages of QGANs in generative chemistry by testing different components of the GAN framework with quantum counterparts. They demonstrated that using a quantum noise generator (prior distribution sampling) could yield compounds with better drug properties. However, they found that quantum generators struggled to generate molecules that resembled those in the training set and encountered computational restrictions during further training. Additionally, they showed that a quantum discriminator with just 50 parameters could achieve a better KL score than a classical discriminator with 22000 parameters, indicating that quantum components can enhance expressive power even with a much fewer number of parameters. Nevertheless, these advancements often compromised the validity and uniqueness of the generated molecules, potentially undermining the efficiency of the sampling and generation processes.

Anoshin et al.~\cite{anoshin_hybrid_2024} introduced a hybrid quantum cycle generative adversarial network for small molecule generation, utilizing the cycle-consistent framework from prior research~\cite{zhu_unpaired_2020, maziarka_mol-cyclegan_2020}. Their approach featured a hybrid generator where a quantum circuit processed the noise vector (prior distribution) and connected to a \revise{MLP} to generate molecular graphs. This method demonstrated comparable, or even improved, performance across various metrics, including uniqueness, validity, diversity, drug-likeness, as well as synthesizability and solubility, highlighting the potential of hybrid quantum-classical architectures in enhancing generative models. However, the study did not provide a detailed comparison of the total number of parameters used, limiting claims about its expressive power.

While QGANs show some promising results in molecular generation, particularly in areas like enhanced drug properties and the potential for better expressive power in discriminators, significant challenges persist. The expressive power derived from full quantum discriminators may come at the cost of compromising other crucial metrics in molecular generation. Additionally, when hybrid networks achieve improvements in drug properties and other metrics, the exact contribution of expressive power offered by the
quantum component becomes less clear. Thus, an outstanding challenge is to achieve enhanced expressive power without sacrificing performance across other critical metrics.

\subsection{Looking Ahead: Quantum Transformers}

Much of the AI revolution is due to the transformer architecture introduced in the ``Attention is All You Need" paper out of Google DeepMind~\cite{vaswani_attention_2023}. This architecture was originally developed for language translation, and consisted of encoder and decoder components which are connected via a cross-attention mechanism. The encoder alone is useful for learning a context-rich representation for a given input sequence by masking some of the sequence and learning to predict the masked parts. The decoder is useful for generating new sequences by learning to predict the next parts of some sequence given a context. Within the realm of biochemistry and drug discovery, transformer encoders have been developed to extract feature vectors from SMILES strings to be used for downstream predictive tasks, and transformer decoders have been used to generate SMILES strings with prespecified characteristics~\cite{wang2021multi, luong_application_2024, shee_directmultistep_2024, shee_site-specific_2024, li_kernel-elastic_2024}. The fundamental capabilities of the transformer architecture are due to the self-attention mechanism where query, key, and value  vectors are computed for each input token (e.g., a sub-word in text or a character in a SMILES strings), attention scores are derived via a scaled dot product of query and key vectors, and softmax normalizes these scores to obtain weights that modulate the aggregation of the value vector, effectively capturing the magnitude with which each token will attend to every other token in the sequence. The self-attention mechanism is executed multiple times in parallel through what is referred to as multi-head attention.

The overwhelming success of the classical transformer in ML has naturally piqued the interest of QML researchers. Most implementations of quantum transformers have been adapted as Vision Transformers (ViTs) rather than for Natural Language Processing (NLP) \cite{xue_end--end_2024,evans_learning_2024, tariq_deep_2024, cherrat_quantum_2024}. While classical ViT models have been utilized in predictive tasks in chemistry and biophysics \cite{khokhlov_image2smiles_2022, ding_efficiently_2024, jha_prediction_2023}, the primary role of transformers in the context of drug discovery has remained with transformer-based generative models.

\begin{figure}[htbp]
  \centering
  \includegraphics[width=1.0\textwidth]{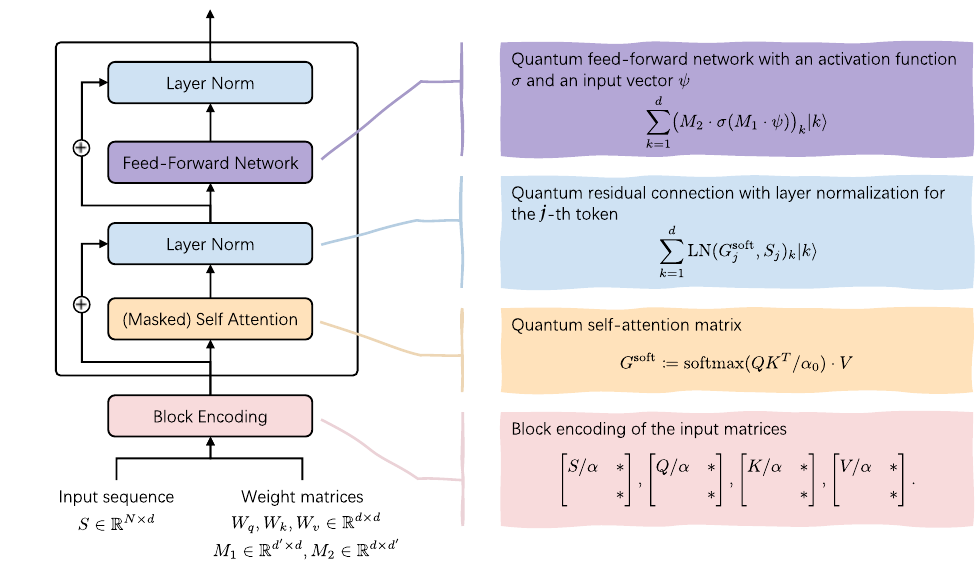}
  \caption{Classical architecture of a transformer decoder layer alongside the corresponding quantum analogs, adapted from Guo \textit{et al.} (2024)\cite{guo_quantum_2024_license}.}
  \label{fig:quantum_transformer}
\end{figure}

Quantum-based attention for generative pre-trained transformers are still in their infancy, and while many of the results presented thus far have been largely theoretical, the field is rapidly advancing. In 2022, DiSipio \textit{et al.}\cite{di_sipio_dawn_2022} discuss the beginnings of quantum NLP, and highlighted that the underlying mathematical operations of the transformer's self-attention mechanism all have implementable quantum formulations. In 2023, both Gao \textit{et al.} \cite{gao_fast_2023} and Li \textit{et al.} \cite{li_quantum_2023} show implementations for a quantum self-attention mechanism. Most recently, Guo \textit{et al.} \cite{guo_quantum_2024_license} and Liao \textit{et al.} \cite{liao_gpt_2024} independently present full end-to-end GPT quantum algorithms. Notably, the work from Guo \textit{et al.} presents a rigorous complexity analysis and demonstrates a theoretical quantum advantage for numerous normalization operations throughout the architecture. The structure of a classical transformer decoder layer and the corresponding quantum implementation by Guo \textit{et al.} is shown in Figure \ref{fig:quantum_transformer}.

The \revise{motivation for} creating a quantum transformer is to reduce the complexity of the self-attention mechanism, which is the bottleneck of the architecture. The traditional classical self-attention mechanism scales $\mathcal{O}(n^2d)$ for sequence length $n$ and embedding dimension $d$. This arises from multiplying the query and key matrices $QK^\top$ as well as applying the resulting pairwise attention matrix to the value matrix $V$. Unfortunately, the current quantum implementations that potentially achieve a complexity advantage rely on assumptions that are seldom true in ML, such as matrix sparsity. Some classical techniques try to avoid the $\mathcal{O}(n^2d)$ complexity of scaled dot-product attention through alternative methods\cite{katharopoulos_transformers_2020,wang_linformer_2020, lee-thorp_fnet_2022}. Similarly - instead of scaled dot-product attention - Quantinuum released an open-source model, Quixer \cite{khatri_quixer_2024}, that proposes a quantum analog of the k-skip-n-gram NLP technique for learning relationships between tokens.
Quixer mixes embedded tokens by using linear combination of unitaries (LCU) \cite{childs_hamiltonian_nodate}, and further computes skip-bigrams between words using quantum singular value transformations (QSVT) \cite{gilyen_quantum_2019}. Quixer's model scales $O(\log(nd))$ in the number of qubits and $O(n\log(d))$ in the number of gates. In contemporary transformer applications, sequence length $n$ is often much larger than the embedding $d$ which makes the logarithmic scaling in the number of qubits with respect to $n$ a promising look into the future of transformers.

While the present models largely do not claim an explicit complexity quantum advantage, this should not dissuade future researchers from utilizing the available methods for their pharmacological applications. The nascency of the field presents an opportunity for researchers in academia and pharmaceutical industry alike to hunt for advantages elsewhere. Presently with no current works in the literature applying quantum transformers to chemical, biological, or pharmaceutical tasks, this should inspire researchers to investigate if these quantum transformers can learn hidden features inaccessible to classical learning styles as indicated by Li \textit{et al.} \cite{li_quantum_2023}. In this event, combining features extracted from both a quantum transformer component and a classical transformer component could present a model with a richer understanding of chemical and biological function, leading to exciting downstream effects in drug design.

\section{Potential of Bosonic Quantum Processors for Quantum Machine Learning}

\subsection{Basics of bosonic quantum computing}

Hybrid qubit-qumode devices \cite{copetudo2024shaping,dutta2024simulating,liu2024hybrid} have the potential to augment the power of qubit architectures by allowing for data encoding in a much larger Hilbert space with hardware efficiency. For example, qumodes could amplify the impact of \revise{VQCs} in applications to \revise{QML} beyond the implementations discussed in \Sec{\ref{sec: variational_quantum_circuits_readout}}.

An arbitrary qumode state $\ket{\psi}$, corresponding to the state of a quantum harmonic oscillator, can be expanded in its Fock basis state representation as a superposition of \revise{a countably} infinite set of orthonormal photon-number states $ \{ \ket{n} \} $.
In practice, however, the expansion is truncated with a Fock cutoff $d$, as follows:
\begin{equation} \label{eq: qumode_fock_basis}
\ket{\psi} 
= \sum_{n = 0}^{d - 1} \: c_n \ket{n}.
\end{equation}
According to \Eq{\ref{eq: qumode_fock_basis}}, a qumode generalizes the two-level qubit into a d-level state (also known as \textit{qudit} \cite{wang2020qudits,roca2024qudit,mandilara2024classification}), thus offering an expanded basis set. Beyond the expanded basis, the hardware of bosonic modes are relatively weakly affected by amplitude damping errors \cite{liu2024hybrid}, which leads to extended lifetimes, and the possibility of implementing efficient error correction codes.~\cite{Ofek2016ExtendLifetime,Sivak2023BreakEven,Ni2023BreakEven}

\begin{figure}[t]
  \centering
  \includegraphics[width=.30\textwidth]{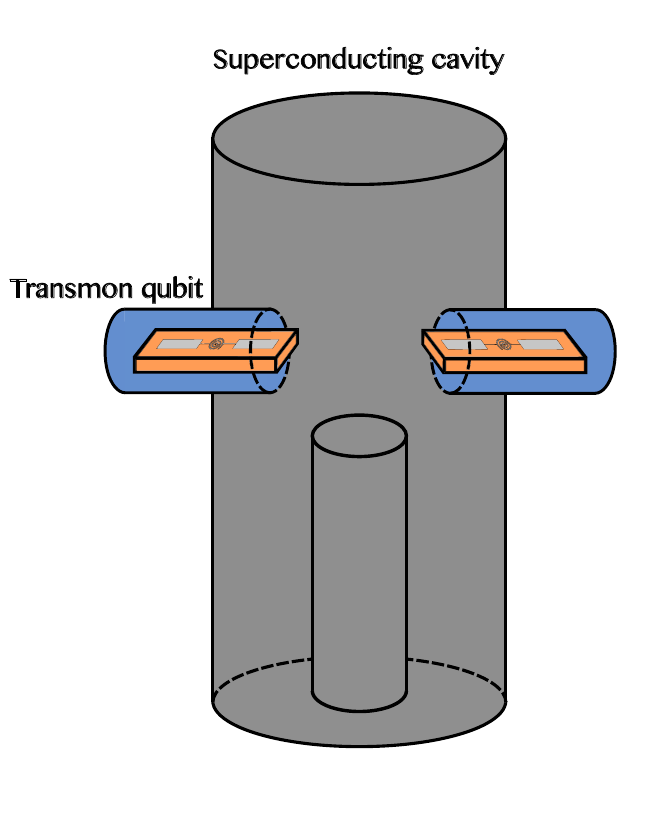}
    \includegraphics[width=1.0\textwidth]{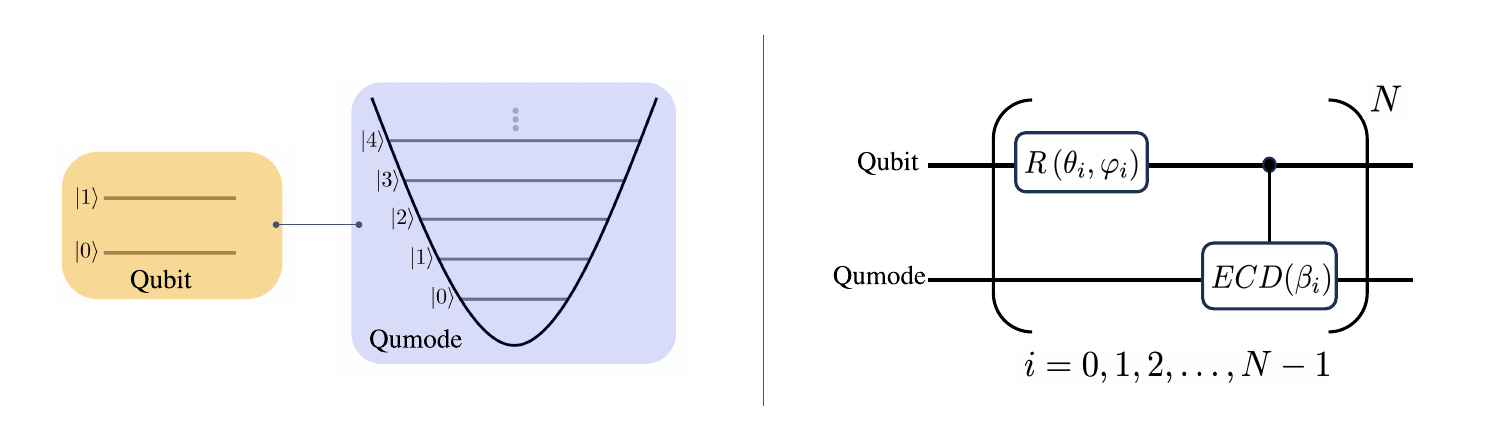}
  \caption{(Top) Schematic representation of a superconducting cavity resonator coupled to a qubit transmon.
  (Bottom Left) Visual schematic of a two level qubit (transmon) coupled to a multi-level oscillator (superconducting cavity). 
  (Bottom Right) Example of a qubit-qumode circuit that allows for universal control, where the qubit rotation gate $R (\theta, \varphi)$ is defined in \Eq{\ref{eq: qubit_rotation}} and the echoed-conditional displacement (ECD) gate is defined in \Eq{\ref{eq: ecd}}. 
  }
  \label{fig:qubit_qumode}
\end{figure}

Recent advancements in bosonic quantum hardware have significantly progressed, enhancing the implementation of qumodes across various architectures \cite{copetudo2024shaping}
However, achieving universal quantum computing remains challenging when relying solely on native qumode operations. This is where hybrid qubit-qumode hardware have made notable strides. For example, in the circuit quantum electrodynamics (cQED) framework, a microwave cavity coupled to a transmon qubit has demonstrated considerable potential (Figure~\ref{fig:qubit_qumode}).~\cite{Blais2021} The interplay between qubit and qumode dynamics enables the development of hybrid qubit-oscillator gate sets, which are efficient in achieving universality. \cite{krastanov2015universal,eickbusch2022fast,liu2024hybrid}. 

Additionally, photonic processors offer programmability that facilitates the simulation of bosonic systems.~\cite{Kalajdzievski2018Hubbard,Zhang2024Hubbard} In contrast, qubit based hardware is inherently suited for simulating fermions through the Jordan-Wigner transformation.~\cite{Bravyi2002Fermion, Ortiz2001Fermion, Ortiz2002Fermion} Therefore, a hybrid qubit-qumode architecture is particularly attractive, particularly since qubit-only or bosonic-only native gates might require deeper circuits for specific applications, although methods have been developed to represent bosons using qubits and vice versa.~\cite{Sawaya2019VibronicSpectra,Shaw2020Schwinger,Rishab2024ElectronicStructure} . 

Incorporating efficient bosonic representation could enable practical simulations beyond the capabilities of conventional qubit-based quantum computers, as already shown for example in \revise{calculations} of vibrational spectra of small polyatomic molecules.~\cite{Wang2020FC} This can be achieved with photonic quantum processors \cite{Huh2015FCBS,Shen2018FCBS,Sparrow2018FCBS}, cQED devices \cite{Wang2020FC}, and even hybrid qudit-boson simulators \cite{MacDonell2023FC}. 

Another unique feature of qumodes is that they can also be represented by continuous variable (CV) bases corresponding to position and momentum operators of a quantum harmonic oscillator, \cite{Weedbrook2012} with no counterpart for qubits. For example, in the position representation, an arbitrary qumode state $\ket{\psi}$ can be expressed, as follows: 
\begin{equation} \label{eq: qumode_position_basis}
\ket{\psi} 
= \int_{-\infty}^{+ \infty}  dx \: \psi (x) \ket{x},
\end{equation}
where $ \psi (x) = \braket{x | \psi} $ is the oscillator complex valued amplitude at $x$. As state and process tomography are necessary to calibrate and model hardware noise, hybrid processors offer simple protocols to determine the Wigner function of qumode states \cite{Home2020Tomography,Bertet2002Tomography,Hofheinz2009Tomography, Vlastakis2015Tomography, Lutterbach1997Tomography}, allowing further development of abstract machine models\cite{liu2024hybrid}.

\subsection{Potential Advantages of Qubit-Qumode Circuits in QML}\label{sec: qubit_qumode_advantage}




Hybrid qubit-qumode circuits, such as the one shown in \Fig{\ref{fig:qubit_qumode}}, can be parameterized with universal ansatzes to approximate any unitary transformation of the qubit-qumode system. 
An attractive choice of a universal ansatz\cite{eickbusch2022fast} applies repeating modules of a qubit rotation gate, 
\begin{equation} \label{eq: qubit_rotation}
R (\theta, \varphi)
= e^{-i\frac{\theta}{2}(\sigma_{x}\cos{\varphi}+\sigma_{y}\sin{\varphi})}
\end{equation}
where $\sigma_{x}$ and $\sigma_{y}$ are Pauli X and Y matrices, followed by an echoed conditional displacement (ECD) gate,
\begin{subequations} \label{eq: ecd}
\begin{align}
ECD (\beta) 
&= \ket{1} \bra{0} \otimes D(\beta/2) 
+ \ket{0}\bra{1} \otimes D(-\beta/2),
\\
D(\alpha) 
&= e^{\alpha \BC{} - \alpha^{*} \BA{}},
\end{align}
\end{subequations}
where $\BC{}$ and $\BA{}$ are bosonic creation and annihilation operators, respectively. 


\revise{
The qumode Hilbert space may offer advantages over qubit-based registers since it allows for more efficient representations for predictive and generative tasks~\cite{wang2020qudits,roca2024qudit,mandilara2024classification}. For example, a system with 8 qubits involves a Hilbert space with $2^{8} = 256$ basis states, which could be represented by two qumodes with control over $d=16$ Fock states (each mode offering a Hilbert space equivalent to the space expanded by 4 qubits)\cite{Wang2020FC}. 
Therefore, encoding of complex molecular information that typically requires many qubits would potentially benefit from hybrid qubit-qumode circuits, as these systems offer significant hardware efficiency compared to qubit circuits with a similarly sized Hilbert space. Additionally, the circuits of qumode states can be based on efficient ansatzes or shallow circuits that bypass the need of deep circuits based on elementary logic gates.\cite{wang2020qudits,Wang2020FC} 
}

\subsection{Encoding Classical Information in Qubit-Qumode Circuits}

We introduce two possible methods for encoding classical (or quantum) data in the form of quantum states of a qumode coupled to a qubit. Similar to amplitude encoding for qubit systems, we can adapt the method discussed in \Sec{\ref{sec: amp_encoding}} for qumodes. We simply modify Eq.~(\ref{amplitude_encoding}) to encode a vector of length $d$ into the amplitudes $\alpha_k$ of a $d$-level qudit, as follows: \begin{equation}
    x \rightarrow U_x\ket{0}_{d} = \ket{x} = \sum_{k=0}^{d-1}\alpha_k\ket{k},
    \label{amplitude_encoding_qumode}
\end{equation}
where $U_{x}$ is the unitary transformation that encodes the data provided by the amplitudes in the form of the qumode state $|x\rangle$. 
Here, $\ket{0}_{d}$ is the initial vacuum state of the qumode corresponding to an empty cavity without photons. Preparing $U_{x}$ requires \revise{parameterization} of an ansatz with universal qumode control such as the one with blocks of a qubit rotation gate followed by an ECD gate (R-ECD ansatz) outlined in Figure \ref{fig:qubit_qumode} and \Sec{\ref{sec: qubit_qumode_advantage}}.\cite{eickbusch2022fast} 
Other ansatzes are also available which can be \revise{parameterized} to encode any arbitrary data set by amplitude encoding in the form of a qumode state.\cite{krastanov2015universal,liu2024hybrid} 

Another practical method for encoding molecular information (e.g., a list of tokens defining a specific molecule) in a qubit-qumode state involves a generalization of phase encoding. A dictionary is used to correlate the input tokens to the values of parameters used in the ansatz. When using the R-ECD ansatz defined by \Eq{\ref{eq: qubit_rotation}} and \Eq{\ref{eq: ecd}}), specific parameters $\theta, \varphi,$ and $\beta$ of each module are assigned to each specific token of the input. So, the sequence of tokens defining the input molecule is encoded as a specific \revise{parameterization} of the R-ECD ansatz. This generalization of phase encoding is not limited to molecular encodings, or the specific choice of ansatz, and can be applied for a wide range of studies, including the implementation of \revise{large language models (LLMs)} in qubit-qumode devices. 

\begin{figure}[htbp]
  \centering
  \begin{subfigure}[b]{0.90\textwidth}
    \centering
    \includegraphics[width=\textwidth]{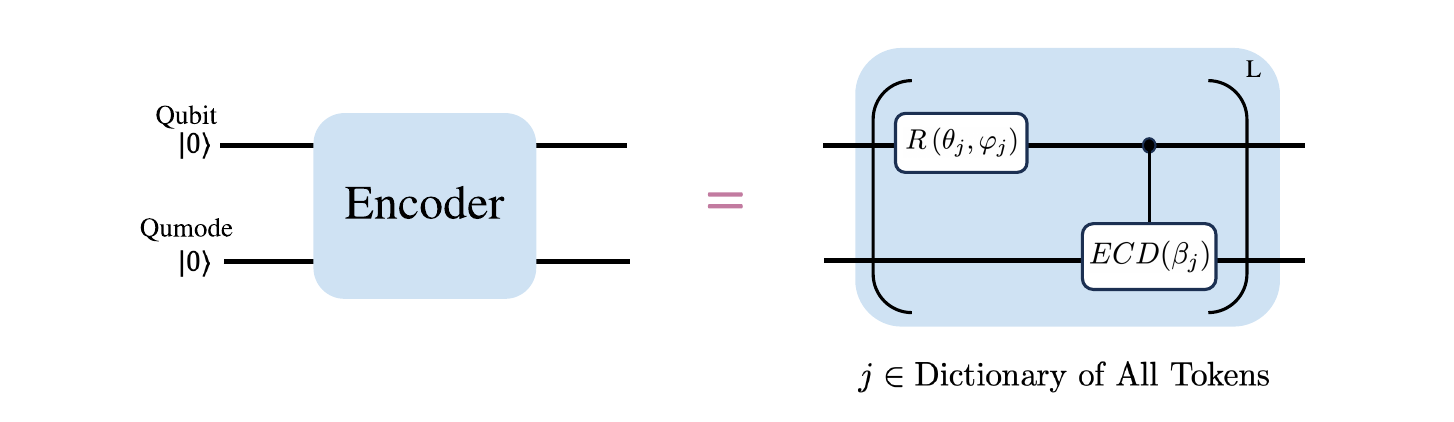}
    \caption{General encoder architecture with the R-ECD circuit.}
    \label{fig:qudit_encode}
  \end{subfigure}
  \hfill
  \begin{subfigure}[b]{0.90\textwidth}
    \centering
    \includegraphics[width=\textwidth]{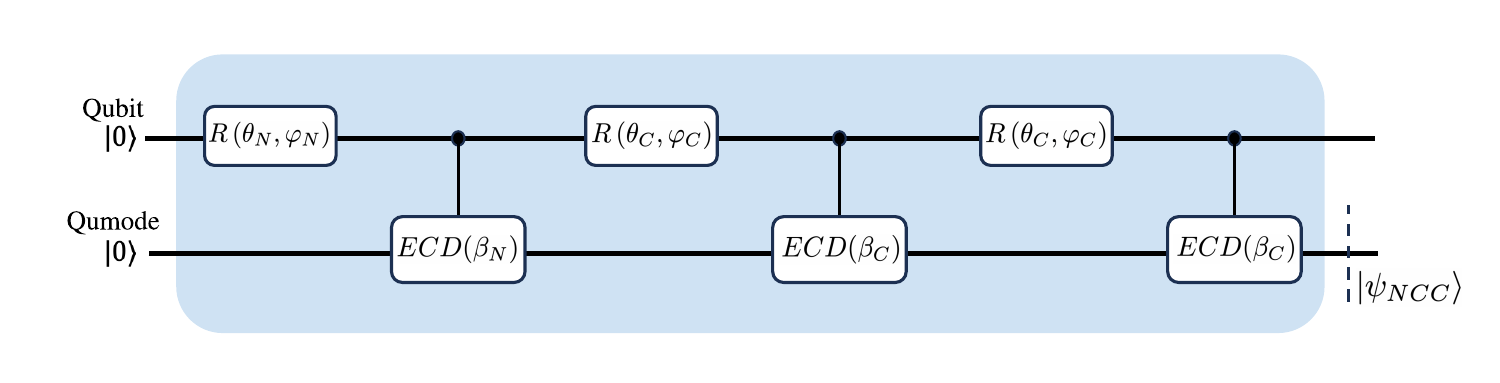}
    \caption{An example showing the encoding circuit of NCC, the SMILES representation of ethylamine. }
    \label{fig:qudit_example}
  \end{subfigure}
  \caption{Circuit diagrams for the R-ECD encoding method. (a) General encoder architecture with a set of $\theta, \varphi$, and $\beta$ assigned to each unique token in the dictionary, encoding a string of length $L$. (b) Specific circuit for the encoding of ethylamine, where the R-ECD block corresponding to $N$ is applied, followed by two blocks of R-ECD corresponding to $C$.}
  \label{fig:RECD_encode}
\end{figure}

One technical challenge of these generalized phase encoding methods is that the encoded states for different states could partially overlap with each other, unless an orthogonalization procedure is enforced. The partial overlap could lead to some level of confusion due to ambiguity of the encoding. To address this challenge, the parameters assigned to each token can be made learnable parameters such that the encodings are optimized to be as different as possible. 

\section{Efficient Circuit Simulation for Near-Term Research and Computing Unit Integration}



\begin{figure}[!htp]
    \centering
    \includegraphics[width=1\textwidth]{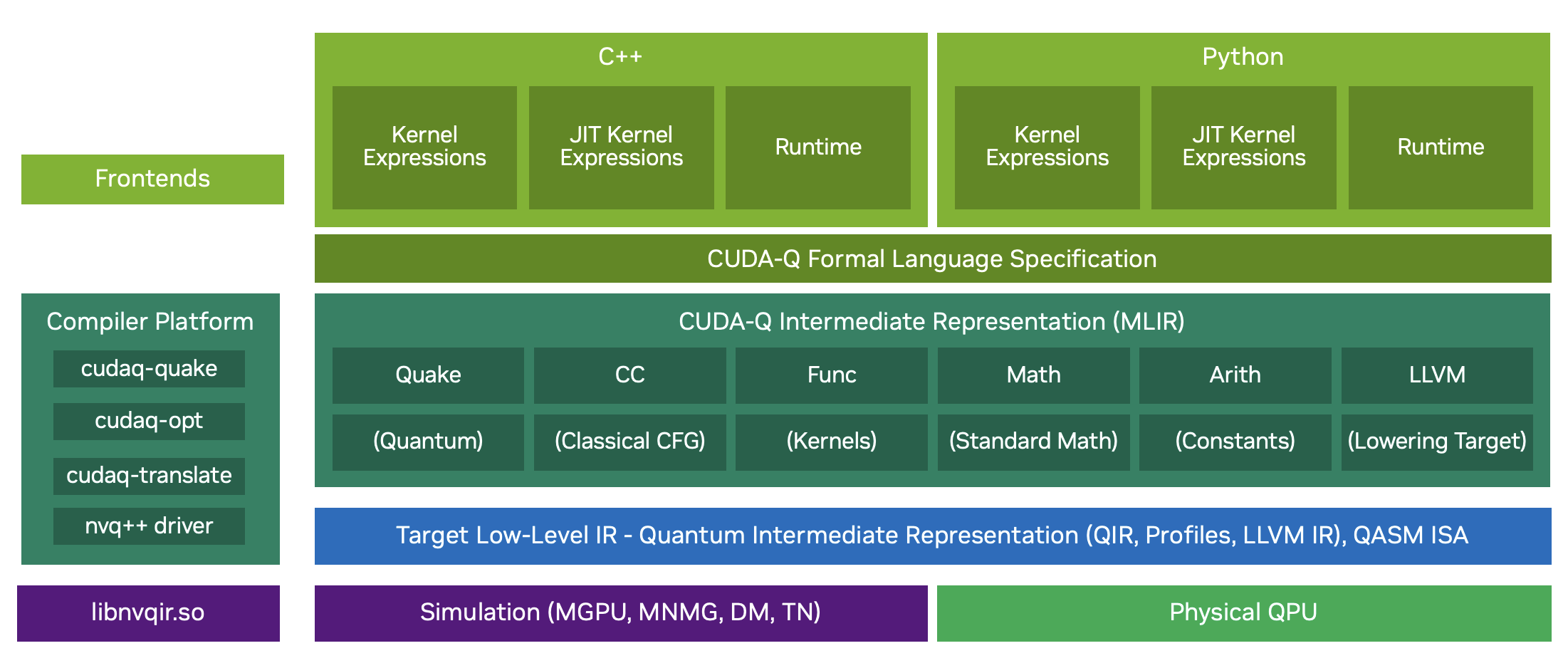}
    \caption{The CUDA-Q software stack. CUDA-Q builds off of a core MLIR-based intermediate representation for representing hybrid quantum-classical code with control flow. The compiler workflow lowers to target specific code for backend QPU execution. This state-of-the-art compiler stack is exposed to programmers via a library-based C++ language extension and a JIT compiled language representation in Python.}
    \label{fig:cudaq-stack}
\end{figure}

\revise{Despite recent progress, current Quantum Processing Units (QPUs) remain limited in size and computational capabilities due to noise and scaling challenges, which impedes progress in algorithmic research. To address this challenge, circuit simulation techniques are meeting the critical need to advance research boundaries.} An open-source platform for seamlessly integrating and programming QPUs, GPUs, and CPUs within a single system is provided by NVIDIA's CUDA-Q \cite{The_CUDA-Q_development_team_CUDA-Q} (see Figure \ref{fig:cudaq-stack}). Various quantum computing frameworks, including Cirq, Qiskit, TorchQuantum, and Pennylane,\cite{bergholm2022pennylaneautomaticdifferentiationhybrid,qiskit2024,cirq-github,hanruiwang2022quantumnas} utilize GPU-accelerated simulation through the cuQuantum libraries \cite{cuquantum-paper} featured in the CUDA-Q simulation backend. By employing the CUDA-Q compiler alongside cuQuantum APIs as simulation backends, users can achieve near-optimal GPU acceleration and exceptional performance at scale.

In this section, we demonstrate how CUDA-Q can be utilized to accelerate and scale up quantum circuit simulations. This is applicable to various fields  including quantum machine learning for chemistry. We use CUDA-Q v0.8 for simulations and show how the compute resources scale with the size of the simulation. Examples used to reproduce the results presented in this section are available in GitHub\cite{GitHub-chem-review}.

\subsection{Circuit simulator with state vector and GPU acceleration.}

\begin{figure}[!htp]
    \centering
    \includegraphics[width=0.6\textwidth]{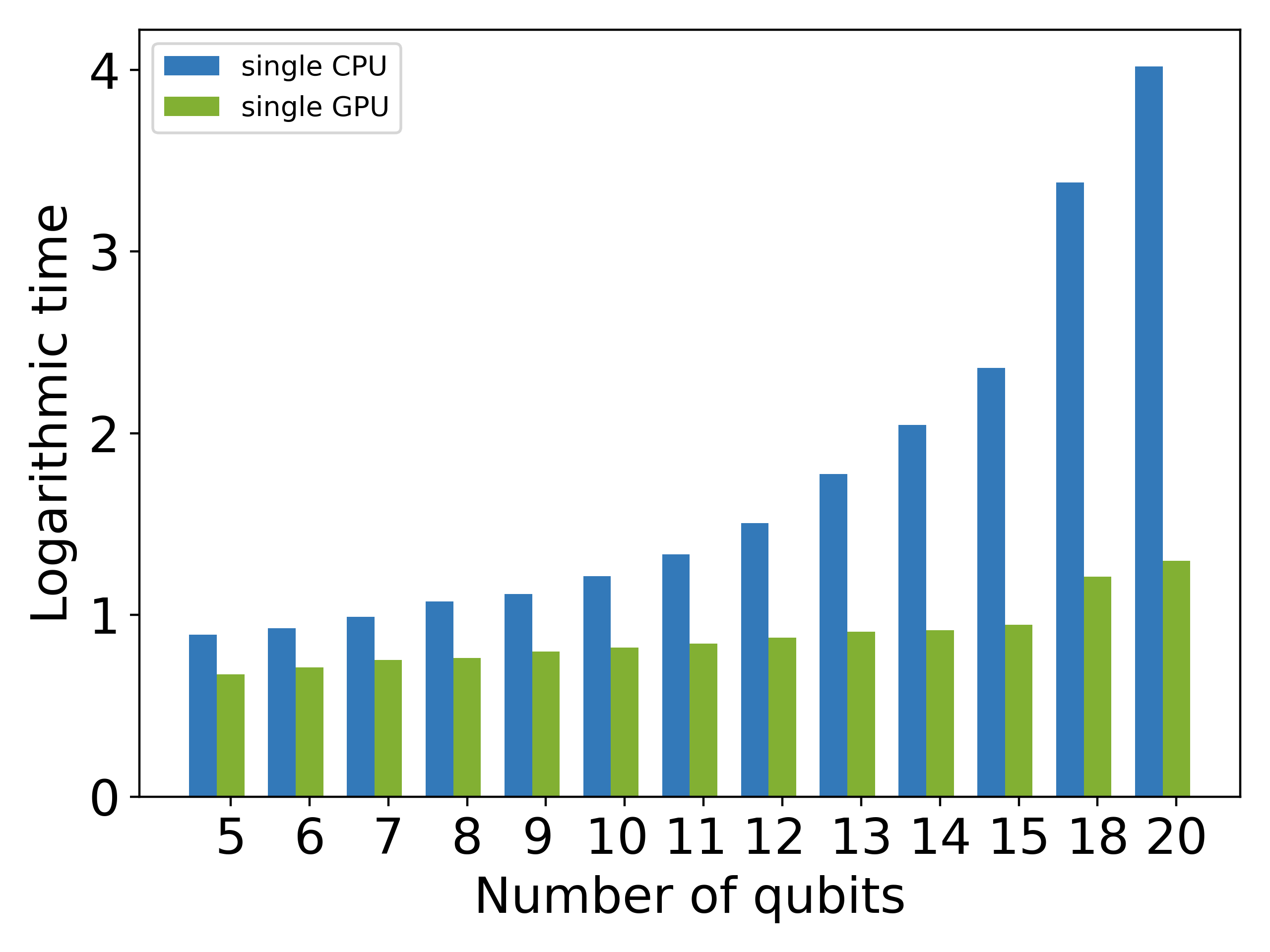}
    \caption{Logarithmic (log$_{10}$) execution time for one `observe' call (i.e., measure the observable operator applied to the state vector/ wave-function, also known as expectation value) for each data-point (10 thousand data-points), i.e., in total there are ten thousand expectation values, on a single CPU versus a single GPU for a one layer of the parameterized quantum circuit (PQC) similar to the PQC shown in Figure \ref{fig:QNN-ex}. The code used to generate the data in this figure is available on GitHub\cite{QNN-1gpu-1cpu}.}
    \label{fig:cpu-gpu}
\end{figure}

Desktop CPUs can handle the simulation of small numbers of qubits; for instance, on a laptop with at least 8 GB of memory, noiseless simulations can reach up to 24 qubits, while noisy simulations are feasible with up to 18 qubits \cite{choose_hw}. However, as the memory required to store the full state vector grows exponentially with the number of qubits, GPUs are needed for larger simulations. For example, an NVIDIA DGX A100 can simulate 20 qubits with exceptional speed, while a CPU would be very slow at performing the state vector simulation of similar size, as shown in Figure \ref{fig:cpu-gpu}.

\begin{figure}[!htp]
    \centering
    \includegraphics[width=0.4\textwidth]{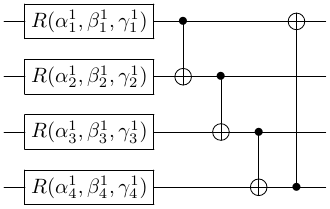}
    \caption{An example of a parameterized quantum circuit employed in QNNs.}
    \label{fig:QNN-ex}
\end{figure}

Figure \ref{fig:cpu-gpu} compares the logarithmic (log$_{10}$) runtime for computing the expectation value of a quantum circuit similar to the one shown in Figure \ref{fig:QNN-ex} using a state vector simulator on one CPU (AMD EPYC 7742 64-Core Processor) as compared to one NVIDIA A100 GPU. The quantum circuit in Figure \ref{fig:QNN-ex} is a standard parameterized quantum circuit employed in QNNs for different applications such as  QGANs applied for drug discovery and \revise{molecular} generation~\cite{kao_exploring_2023, IEE_QGAN_Alexey_2024}. Specifically, Figure \ref{fig:cpu-gpu} shows the comparison of the runtime on a single CPU as compared to a single GPU for data-points of ten thousand (i.e., ten thousand expectation values) as a function of the number of qubits. It is shown that the runtime on the CPU significantly increases as we increase the number of qubits while increases only modestly on the NVIDIA A100 GPU. For example, for the 18 qubit circuit, there is a $\approx$150$\times$ speed up on the single GPU. When increasing the number of qubits to 20, the speed up is $\approx$530$\times$. These results emphasize the need for GPU supercomputing to accelerate simulations of quantum algorithms and applications to research and development. Such simulations would enable studies beyond small-scale proof-of-concept calculations in application studies to real-world scenarios.


Another example demonstrating the capabilities of CUDA-Q are the implementations of the VQE-\revise{Quantum approximate optimization algorithm (VQE-QAOA)} algorithm for simulations of molecular docking.~\cite{PhysRevApplied.21.034036} and protein folding\cite{prtein-folding-nature-2021} For example, the VQE-QAOA algorithm has been applied to find the optimal configuration of a ligand bound to a protein, implementing the molecular docking simulation as a weighted maximum clique problem.~\cite{PhysRevApplied.21.034036} Simulations were performed with up to 12 qubits. The CUDA-Q tutorial \cite{QAOA-mdocking} reproduces the results using DC-QAOA and compares the CPU and GPU runtimes (Table \ref{dcqaoa-mdoc}). For 12 qubits, a 16.6$\times$ speed up is observed on a single GPU when compared to a single CPU. 
\begin{table}
\caption{Execution time of one `observe' call (i.e., expectation value) using DC-QAOA ansatz. Simulations were run with 3, 8, and 13 layers for 6, 8, and 12 qubits, respectively.}
\label{dcqaoa-mdoc}
\begin{tabular}{ccc}
\hline
Qubits & CPU time (s) & GPU time (s)\\
\hline
6   & 0.322 & 0.160\\
8   &1.398  &0.390 \\
12  &6.863  &0.412 \\
\hline
\end{tabular}
\end{table}

\begin{figure}[!htp]
    \centering
    \includegraphics[width=0.7\textwidth]{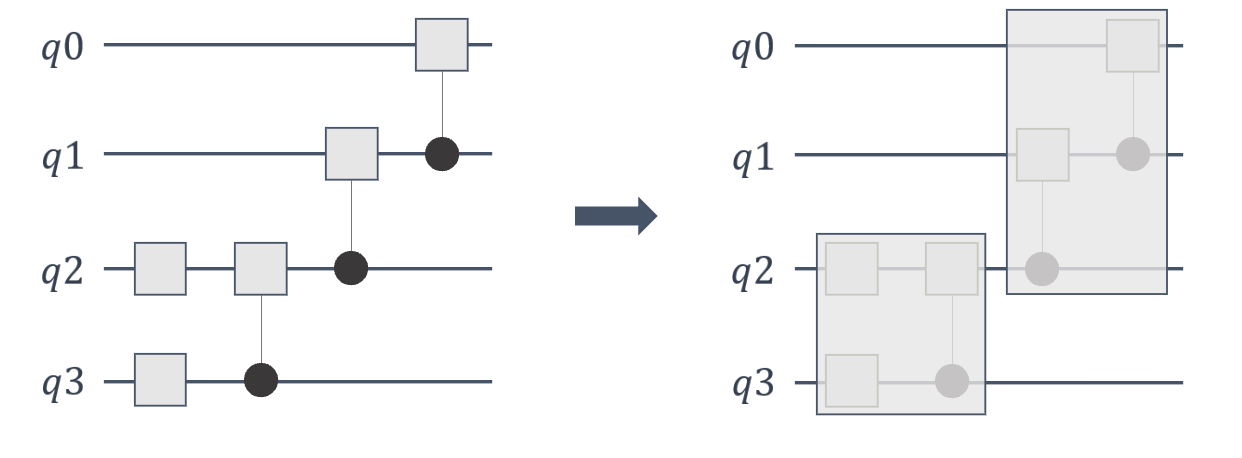}
    \caption{Gate fusion fuses multiple gates into one larger gate.}
    \label{fig:gate-fuse}
\end{figure}

CUDA-Q also allows for gate fusion to enhance state vector simulations with deep circuits, thereby improving performance.~\cite{gate_fusion_blog, alan-qHiPSTER} Gate fusion is an optimization technique that combines consecutive quantum gates into a single gate (see Figure \ref{fig:gate-fuse}), which reduces the overall computational cost and increases the circuit efficiency.~\cite{gate_fusion_doc, gate_fusion_cudaq} By grouping small gate matrices into a single multi-qubit gate matrix, the fused gate can be applied in one operation, eliminating the need for multiple applications of small gate matrices. This optimization reduces memory bandwidth usage, as applying a gate matrix G to a state $|\Psi\rangle = G  |\phi \rangle$ involves reading and writing the state vector. The memory bandwidth (in bytes, including reads and writes) can be calculated, as follows:
\begin{equation}
\mathrm{memory\,\,\, bandwidth} = 2 \,\times  \frac{\mathrm{svSizeBytes}}{2^{\mathrm{ncontrols}}},
\end{equation}
where `svSizeBytes' represents the state vector size in bytes and `ncontrols' is the number of control \revise{qubits} (e.g., a CNOT gate has one control). Applying two gates, $G_2 G_1 |\phi\rangle$, requires two reads and two writes, whereas applying the combined gate $(G_1 G_2)|\phi\rangle$ only needs one read and one write.

Gate fusion can significantly enhance simulation performance for deep circuits which are crucial for quantum applications in chemistry. A notable example is the unitary coupled cluster singles and doubles (UCCSD) ansatz, widely used in quantum \revise{computational} chemistry calculations. For instance, when running a single observation call ({\em i.e.}, computing one expectation value) for the C$_2$H$_4$ molecule using the UCCSD ansatz with 24 qubits on an NVIDIA A100, the total elapsed time is 30.02 second without gate fusion. In contrast, with gate fusion, the elapsed time is reduced to 12.44 second, demonstrating a 2.4$\times$ speedup. The code for this comparison is available on GitHub\cite{gate-fuse-ex}.

\subsection{Parallelization and Scaling}
NVIDIA's CUDA-Q platform provides a clear overview of the various devices in a quantum-classical compute node, including GPUs, CPUs, and QPUs. Researchers and application developers can work with a diverse array of these devices. Although the integration of multiple QPUs into a single supercomputer is still in progress, the current availability of GPU-based circuit simulators on NVIDIA multi-GPU architectures enables the programming of multi-QPU systems today. 

\subsubsection{Enabling Multi-QPU Workflows}

CUDA-Q enables application developers to design workflows for multi-QPU architectures that utilize multiple GPUs. This can be achieved using either the `NVIDIA-mQPU' backend~\cite{cudaq-backend} or the `remote-mQPU' backend, which we discuss further in Sec.~\ref{remote-mqpu-gpu}. The `NVIDIA-mQPU' backend simulates a QPU for each available NVIDIA GPU on the system, allowing researchers to run quantum circuits in parallel and thus accelerating simulations. This capability is crucial for applications such as quantum machine learning algorithms. For example, in training QNNs, computing expectation values for numerous data-points is often \revise{required} to train the model. By batching these data-points, they can be processed simultaneously across multiple GPUs. 

\begin{figure}[!htp]
    \centering
    \includegraphics[width=0.5\textwidth]{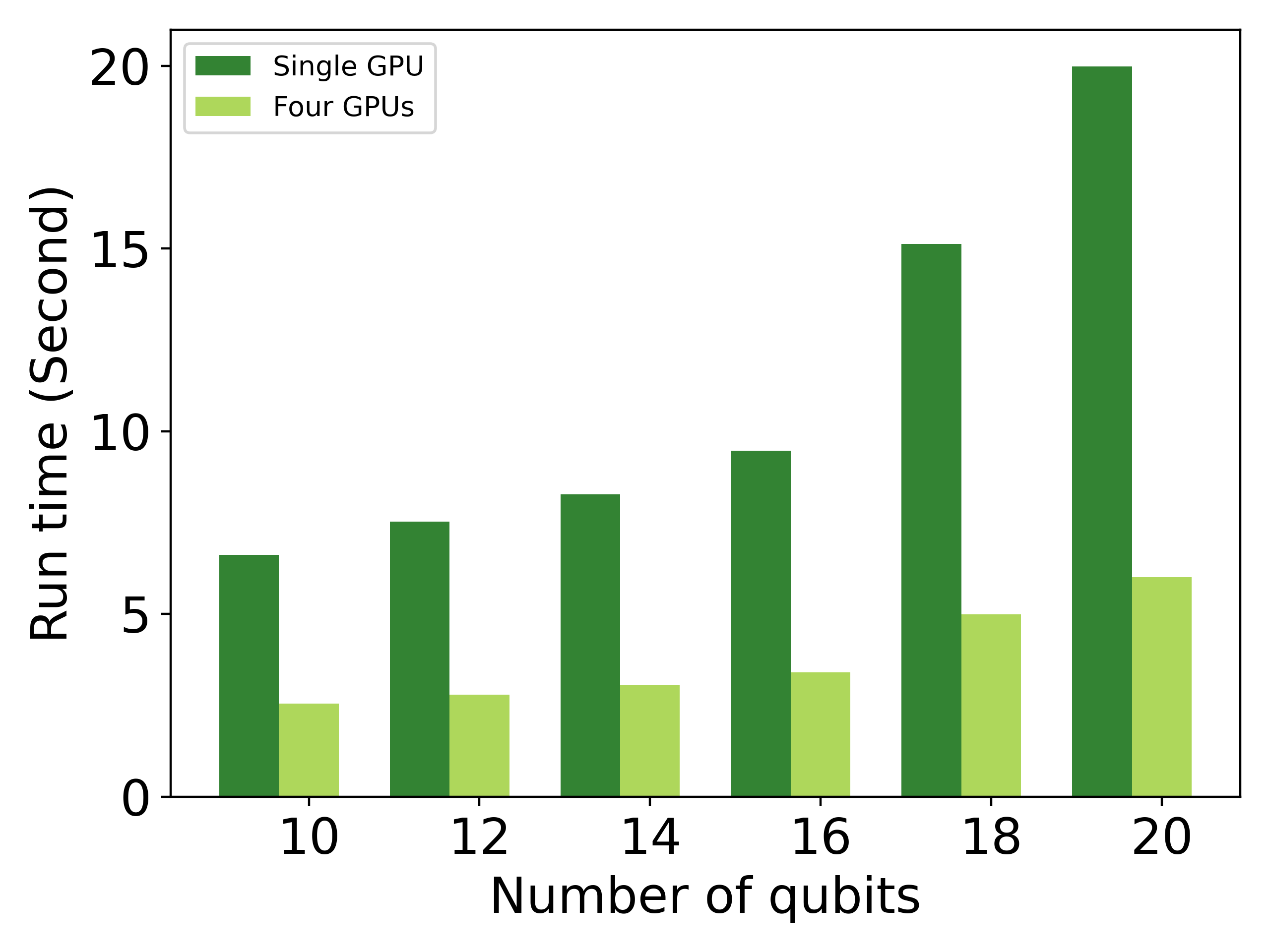}
    \caption{Execution time for `observe' call (expectation value) made for ten thousand data-points for a one layer of the parameterized quantum circuit similar to the one shown in Figure \ref{fig:QNN-ex}. All simulations were run on NVIDIA DGX A100 GPU device. For a single GPU, the total ten thousand data-points are dealt within a single GPU (i.e, ten thousand `observe' call in sequential on a single GPU). For the four GPU case, the data-points are split into four batches, each containing 2500 data-points (i.e., 2500 `observe' calls on each GPU).}
    \label{fig:1gpu-4gpu}
\end{figure}

Figure \ref{fig:1gpu-4gpu} compares results obtained by running a QNN workflow running on a single GPU versus those obtain by distributing the workflow across four GPUs (in a single CPU node with 4 GPUs). The code for this comparison is available in GitHub\cite{QNN-mQPU}. For an application using 20 qubits, we find that the runtime with four-GPUs is approximately 3.3 times faster than using a single GPU. Although parallelization requires some synchronization and communication across the GPUs, which slightly limits the speedup to being less than 4x, this still demonstrates strong scaling performance. It highlights the efficient utilization of GPU resources when available.

Another example of a commonly used application primitive that benefits from parallelization using the `NVIDIA-mQPU' backend is the Hadamard test. The Hadamard test is crucial for computing the overlap between different states, as necessary to evaluate correlation functions, and expectation values which involve calculating $O(n^2)$ independent circuits in a wide range of applications, including prediction of drug toxicity~\cite{smaldone_quantum--classical_2024} and determining the electronic ground state energy of molecules~\cite{MRQKS-U0Emory,PRXQuantum.2.040352}. By leveraging parallelism, these $O(n^2)$ circuits can be efficiently executed across as many QPUs --whether physical or simulated-- as are available.

\subsubsection{Scaling Circuit Simulations with Multi-GPUs}\label{scaling}

The conventional state-vector simulation method requires storing 2$^n$ complex amplitudes in memory when simulating $n$ qubits. This results in exponentially increasing memory requirements for circuits with a large number of qubits. If each complex amplitude requires 8 bytes of memory, the total memory required for an n qubit quantum state is 8 bytes $\times$ 2$^n$. For instance, with $n=30$ qubits, the memory requirement is approximately $8$ GB, while for $n=40$ qubits, it jumps to about $8700$ GB. CUDA-Q addresses this challenge by enabling the distribution of state-vector simulation across multiple GPUs via the `NVIDIA-mGPU' backend.~\cite{cudaq-backend} For detailed information of the algorithm, see Sec. II-C in Ref. \citenum{cuquantum-paper}. Additionally, examples of using the `NVIDIA-mGPU' backend are available on GitHub\cite{ghz-mGPU}.

The `NVIDIA-mGPU' backend combines the memory of multiple GPUs within a single DGX compute node and across multiple DGX compute nodes in a cluster. DGX compute nodes, part of NVIDIA’s DGX platform, are high-performance computing (HPC) servers, specifically designed for HPC and artificial intelligence (AI) workloads, leveraging NVIDIA GPUs to accelerate intensive computations. By pooling GPU memory, this backend allows for greater scalability and eliminate the memory limitations of individual GPUs. Consequently, the capacity to simulate larger numbers of qubits is constrained only by the available GPU resources in the system.

Intra-node NVlink~\cite{nvlink} is a powerful tool for large-scale simulations. An NVLink-based system enables greater performance optimization by providing direct access to the full NVLink feature set, bypassing the CUDA-Aware MPI layer. CUDA-Q v0.8 introduces an improved algorithm for intra-node NVLink, leveraging CUDA Peer-to-peer (P2P) communication.~\cite{CUDA-p2p} Table \ref{table:nvlink} compares the performance of CUDA-Q 0.7 (using CUDA-aware MPI) and CUDA-Q 0.8 (using P2P) on an NVlink-enabled DGX H100 system. In these simulations, the state vector was distributed across a single node with 8 GPUs. Four large-scale quantum algorithms were benchmarked using the MPI and P2P API in CUDA Runtime. As shown in Table \ref{table:nvlink}, CUDA-Q v0.8 with P2P achieves up to 2.5x speedup for H-gates compared to CUDA-Q v0.7 with CUDA-aware MPI. 

\begin{table}
\caption{Quantum algorithm performance improvements enabled by NVLink optimizations. The speed up in time is reported for CUDA-Q v0.8 with CUDA P2P compared to CUDA-Q v0.7 with CUDA-aware MPI. Simulation times accounts for a single VQE execution. H-Gates refers to applying one Hadamard gate per qubit. All simulations were run on a DGX H100 device.}
\label{table:nvlink}
\begin{tabular}{ccc}
\hline
Algorithm& Qubits&Speed up (in simulation time)\\
\hline
H-Gates & 35 & 2.47\\
QAOA & 32 & 1.28\\
QFT & 35 & 1.13\\
UCCSD &32&1.30\\\hline
\end{tabular}
\end{table}

Additionally, developers can now use CUDA-Q to fully exploit the performance of the NVIDIA GH200 AI superchip,~\cite{GH200-superchip} further enhancing the capabilities of quantum simulation in CUDA-Q. With a combined CPU and GPU memory of 1.2TB, the GH200 AI superchip significantly accelerates quantum simulations, reducing the number of required nodes by 75$\%$. This reduction is particularly crucial for quantum applications research, which is often constrained by memory limitations.

Table \ref{table:GH200} compares the performance of the GH200 superchip and the DGX H100 for running a quantum algorithm using a state vector simulator. In this comparison, we employed 37 qubits and distributed the state vector across 8 GPUs on four nodes in the GH200 superchip and a single node in the DGX-H100. Our findings show that the GH200 superchip achieves up to 2.58x speed up for the quantum Fourier transform (QFT) and a 4x speed up for H-Gates.

\begin{table}
\caption{Comparison of performance of the GH200 superchip and the DGX H100 for running a quantum algorithm using a state vector simulation. Simulations were run with 37 qubits and the state vector was distributed across 8 GPUs in a single node in the DGX H100 and four nodes in the GH200 superchip. H-Gates refers to applying one Hadamard gate per qubit.}
\label{table:GH200}
\begin{tabular}{ccc}
\hline
Algorithm& Qubits&Speed up (in simulation time)\\
\hline
H-Gate & 37 & 4.10\\
QFT & 37& 2.58\\\hline
\end{tabular}   
\end{table}

\subsubsection{Combining Backends For Large Scale Simulations.}\label{remote-mqpu-gpu}

\begin{figure}[!htp]
    \centering
    \includegraphics[width=0.6\textwidth]{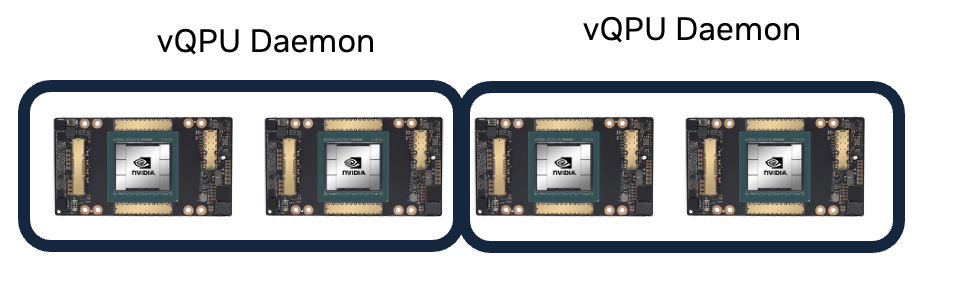}
    \caption{An example of multi-QPU backend with multi-GPU. Here, there are two virtual QPUs (vQPU) and each virtual QPU is made of two GPUs.}
    \label{fig:remote-mQPU}
\end{figure}

Quantum circuit simulations can be scaled up using the `NVIDIA-mGPU' backend and parallelized with the `NVIDIA-mQPU' backend, as described in the previous section. CUDA-Q provides the capability to combine both backends through the `remote-mQPU' backend, enabling large-scale simulations (Figure \ref{fig:remote-mQPU}). In this configuration, \revise{multiple GPUs comprise a virtual QPU}. A practical example of using `remote-mQPU' for QNNs is available on GitHub\cite{remote-mQPU}.   

\subsection{Quantum Circuit Simulator With Tensor Networks}

\begin{figure}[!htp]
    \centering
    \includegraphics[width=0.5\textwidth]{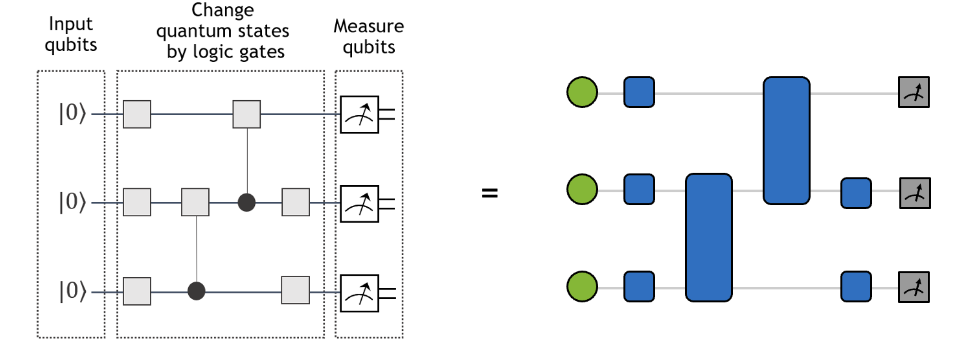}
    \caption{Single-qubit and two-qubit gates translate to rank-2 and rank-4 tensors, respectively. The initial single-qubit states $|0\rangle$ and single-qubit measurement operations can be viewed as vectors (projectors) of size 2. The contraction of the tensor network on the right yields the wavefunction amplitude of the quantum circuit on the left for a particular basis state.}
    \label{fig:tensornet-cudaq}
\end{figure}

The state vector method is effective for simulating deep quantum circuits, however, it becomes \revise{impractical} for simulations of circuits with large numbers of qubits due to the exponential growth in computational resources required --making them unmanageable even on the most powerful supercomputers available today. As an alternative, the tensor network method represents the quantum state of $N$ qubits through a series of tensor contractions (see Figure~\ref{fig:tensornet-cudaq}). This approach allows quantum circuit simulators to efficiently handle circuits with many qubits.

\begin{figure}[!htp]
    \centering
    \includegraphics[width=0.5\textwidth]{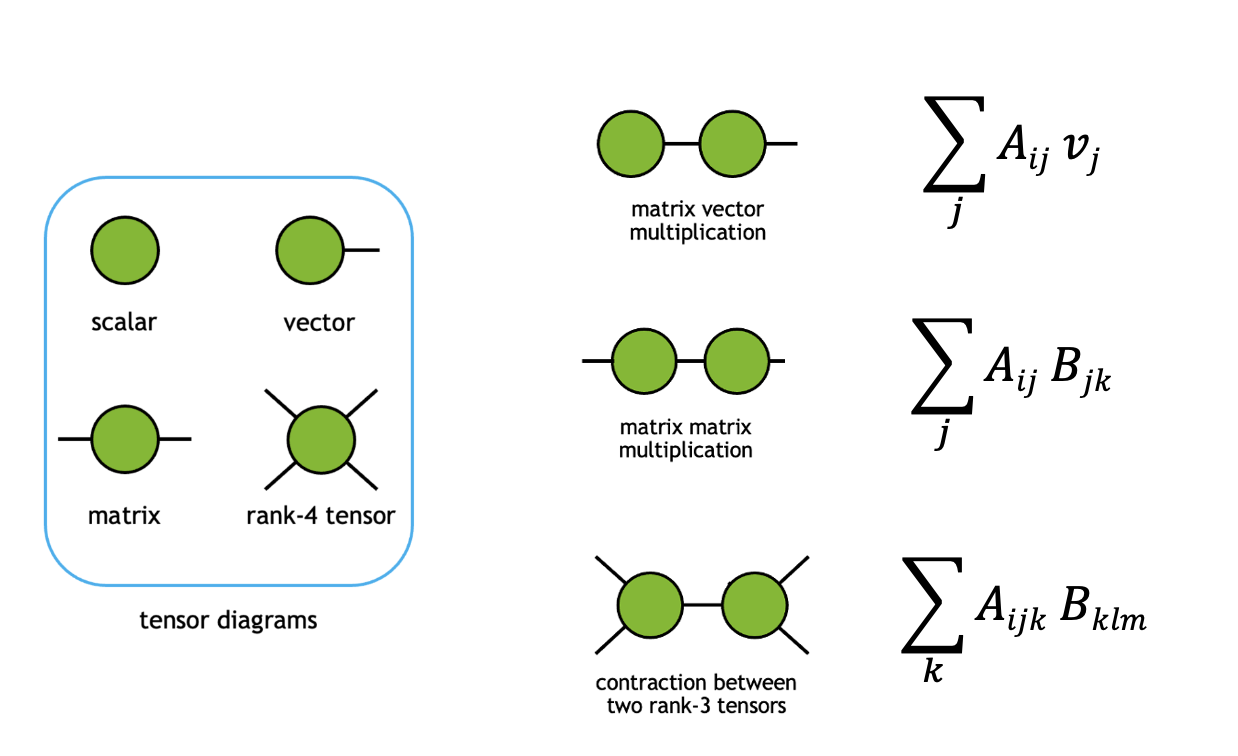}
    \caption{Tensor diagram (left) and example of matrix like contractions (right).}
    \label{fig:tensors-ex}
\end{figure}

Tensors (see Figure \ref{fig:tensors-ex}) generalize scalars (0D), vectors (1D), and matrices (2D) to an arbitrary number of dimensions. A tensor network consists of a set of tensors connected together through tensor contractions to form an output tensor. In Einstein summation notation, a tensor contraction involves summing over pairs of repeated indices (see Figure \ref{fig:tensors-ex}). For example, a rank-four tensor $M$ can be formed by contracting two rank-three tensors $C$ and $B$, as follows: $M_{ijlm}=\sum_k C_{ijk}\, B_{klm}$. Here, the contraction is performed by summing over the shared index $k$. Identifying an efficient contraction sequence is \revise{essential for minimizing the computational cost of the tensor networks}.~\cite{cuquantum-paper,optTN-quantum-2021-Gray} The contractions between the constituent tensors define the topology of the network.~\cite{Or_s_2014,trus-TN-2019}

CUDA-Q offers two GPU-accelerated tensor network backends: `tensornet' and `tensornet-mps'\cite{tensornet-cudaq}. For a detailed explanation of tensor network algorithms and their performance, see Refs.~\citenum{cutensornet, cuquantum-paper}. 

The `tensornet' backend represents quantum states and circuits as tensor networks without any approximations. It computes measurement samples and expectation values through tensor network contractions.~\cite{tensornet-cuquantum} This backend supports the distribution of tensor operations across multiple nodes and GPUs, enabling efficient evaluation and simulation of quantum circuits.

The `tensornet-mps' backend utilizes the matrix product state (MPS) representation of the state vector, exploiting low-rank approximations of the tensor network through decomposition techniques such as QR and singular value decomposition. As an approximate simulator, it allows truncation of the number of singular values to keep the MPS size manageable. The `tensornet-mps' backend supports only single-GPU simulations. Its approximate nature enables it to handle a large number of qubits for certain classes of quantum circuits while maintaining a relatively \revise{low} memory footprint.

\section{Challenges and Outlook} \label{challenges}

\subsection{Hardware}

When evaluating the physical implementation of quantum computers, it is essential to consider the widely recognized five criteria proposed by DiVincenzo~\cite{divincenzo2000physical}:
\begin{enumerate}[label=(\arabic*)]
    \item \textbf{Scalable physical systems with well-characterized qubits}: The system should contain qubits that are not only distinguishable from each other but also manipulable either individually or collectively. This requirement ensures that qubits can be controlled with precision for complex quantum computations.
    
    \item \textbf{Ability to initialize qubits to a simple, known state}: Typically referred to as a ``fiducial" state, this criterion emphasizes the importance of preparing qubits in a well-defined, simple initial state, such as the zero state. This initialization process is crucial for the reliability and predictability of subsequent quantum operations.
    
    \item \textbf{Decoherence times much longer than gate operation times}: Quantum systems must exhibit long coherence times relative to the time it takes to perform quantum gate operations. This ensures that quantum information is preserved long enough to complete computations before being lost to decoherence.
    
    \item \textbf{A universal set of quantum gates}: The hardware must support a set of quantum gates capable of performing any quantum computation. This typically includes a variety of single-qubit gates along with a two-qubit entangling gate, such as the CNOT gate, enabling the construction of complex quantum circuits.
    
    \item \textbf{Qubit-specific measurement capability}: The system should allow for accurate measurement of individual qubits' states after computation. This criterion is essential for retrieving the final output of quantum computations.
\end{enumerate}

Gate-based quantum computer designs generally adhere to these criteria, yet achieving the most optimal performance remains a significant challenge. For QML, these hardware requirements introduce additional complexities.

QNNs often claim superior expressive power compared to classical neural networks. This advantage typically necessitates high connectivity among qubits, aligning with the need for well-characterized and scalable qubit systems described in Criterion (1). Ensuring such connectivity while maintaining system scalability and qubit fidelity is a non-trivial challenge in current hardware implementations.

Moreover, QML algorithms frequently utilize amplitude encoding, a technique that effectively encodes classical data into quantum states. This approach, however, is equivalent to preparing arbitrary quantum states, which goes beyond the simpler requirement of initializing qubits to a fiducial state as outlined in Criterion (2). Consequently, specific QML applications may require either modifications to the existing hardware criteria or the development of more advanced state preparation algorithms to achieve the desired outcomes.

Finally, when the final output of a QNN necessitates precise amplitude measurements of quantum states, the hardware must extend the measurement capabilities described in Criterion (5). Specifically, accurate and scalable quantum state tomography becomes essential to extract the necessary information from the quantum system. This represents another area where current quantum hardware may need further refinement to fully support the demands of QML.

\subsection{Algorithms} \label{challenges_algorithms}

Loading classical data into a quantum state is the often first step in a QNN, and is a step that will largely dictate the performance of the model, the potential advantages the quantum model possess over the classical, and the model's quantum resource complexity. For example, angle encoding \ref{angle_encoding} is inexpensive to implement on quantum hardware, but it is difficult to extract a complexity advantage. Alternatively, amplitude encoding easily enables a complexity advantage due to the exponentially larger Hilbert space in which information can be stored, but at the expense of quantum resources to prepare such quantum state. In particular, state preparation techniques to prepare arbitrary state vectors scale exponentially with respect to the number of CNOT gates required to prepare the quantum state \cite{shende_synthesis_2006, mottonen_transformation_2004}. While this problem of state preparation may be daunting, promising data encoding workarounds are being developed. Data re-uploading is a strategy that allows circuits to handle more complex data by breaking the information into smaller quantum circuits \cite{perez-salinas_data_2020}. Shin \textit{et al.} presents a method for QML that utilizes a quantum Fourier-featured linear model to exponentially encode data in a hardware efficient manner \cite{shin_exponential_2023}. The authors demonstrate the method achieves high expressivity and exhibits better learning performance compared to data re-uploading, notably when learning the potential energy surface of ethanol. These promising directions should motivate QML researchers to identify tasks where their input data exists in or can be transformed into a form that is known to be efficiently prepared\cite{zylberman_efficient_2024,shukla_efficient_2024,gleinig_efficient_2021} or where the exact input vector does not need to be known a priori and is learned through training. Furthermore, as QPUs evolve to include more qubits and improved interconnected topologies, state preparation algorithms that utilize ancillary qubits will help address the challenges associated with poor decoherence times and prolonged gate execution times, as they are capable of preparing arbitrary states with shallower depths \cite{zhang_circuit_2024}.

Similar to how classical ML architectures have the potential to suffer from vanishing gradients, VQCs have the potential to suffer from barren plateaus. Barren plateaus occur when the loss differences used to compute quantum weight gradients exponentially vanish with the size of the system. Larocca \textit{et al.} present comprehensive review where the authors outline strategies to avoid and mitigate the problem of barren plateaus \cite{larocca_review_2024}. Some of these methods the aspiring QML researcher should be aware of are shallow circuits and clever weight initialization strategies. Notably, Ragone \textit{et al.} \cite{ragone_lie_2024} present a theorem to determine exactly if any \revise{noiseless} quantum circuit will exhibit barren plateaus regardless of the circuit's structure. The authors note that among the implications of their work, it is possible to design variational quantum circuits that exhibit high entanglement and use non-local measurements while still avoiding barren \revise{plateaus}, going against conventional wisdom. This lifts restrictions and gives researchers a much deeper insight into the trainability of their circuits.

In addition to the difficulties of determining quantum gradients, updating the quantum weights can prove difficult as well. Classical neural networks have had tremendous success using backpropagation to update the model's weights, however methods for updating quantum weights is still being intensely researched. QNNs most commonly employ the parameter-shift method \cite{mitarai_quantum_2018,schuld_evaluating_2019} to estimate quantum gradients for each weight, however this can prove expensive as it requires running at least $2M$ quantum circuits for $M$ trainable parameters during the backwards pass computation, giving a total time complexity of $O(M^2)$. New methods for quantum backpropagation are emerging that is making the evaluation of quantum gradients more efficient, most recently the work by Abbas \textit{et al.} \cite{abbas_quantum_2023} that reduces the complexity from quadratic parameter-shift method to $O(M\text{polylog}(M))$ time. The expensive nature of required quantum resources to update weights encourage many to optimization methods. Many quantum neural networks in the literature often employ the Constrained Optimization by Linear Approximations algorithm \cite{powell_direct_1994} for weight optimization, however this method is only applicable for models with few trainable parameters.  Work is being done to improve gradient-free based optimization of VQC parameters that are more efficient than the parameter-shirt method. Kulshrestha \textit{et al.} devise an optimization scheme with good scalability potential that trains at the level of classical optimizers while outperforming them in computation time \cite{kulshrestha_learning_2023}. Weidmann \textit{et al.} present an optimization method that significantly improves convergence of QNNs compared to the parameter-shift method \cite{wiedmann_empirical_2023}.

\subsection{Outlook}
In this review, we have examined the use of \revise{QNNs} implemented on gate-based quantum computers for applications in chemistry and pharmaceuticals. While the integration of quantum computing into these fields holds the potential for significant advancements, it also presents unique challenges that must be addressed.

As discussed in the previous subsections, the hardware and algorithmic challenges for QML are substantial. The requirements for coherence, qubit connectivity, and state preparation introduce significant hurdles that have yet to be fully overcome. QNNs often require precise qubit control and extended coherence times, which current quantum hardware struggles to provide consistently. On the algorithmic front, issues such as state preparation, barren plateaus, and efficient quantum gradient computation remain critical bottlenecks that demand innovative solutions.

Recent progress in quantum error correction, highlighted by Google Quantum AI’s breakthrough~\cite{acharya2024quantum}, marks a significant milestone. This achievement suggests that we are nearing the development of more reliable quantum systems, which is crucial for the practical implementation of QML in real-world scenarios. However, there remains a pressing need for improved scalability of quantum hardware and the development of more robust error correction protocols.

Looking ahead, as quantum technology continues to mature, we anticipate the emergence of more sophisticated applications, such as the discovery of new drugs and materials, the optimization of chemical reactions, and the exploration of molecular structures with unprecedented accuracy. The intersection of quantum computing and machine learning offers a unique opportunity to transform how we tackle some of the most complex challenges in science and industry.

\begin{acknowledgement}
The authors acknowledge support from the National Science Foundation Engines Development Award: Advancing Quantum Technologies (CT) under Award Number 2302908. VSB also acknowledges partial support from the National Science Foundation Center for Quantum Dynamics on Modular Quantum Devices (CQD-MQD) under Award Number 2124511.
\end{acknowledgement}

\section*{Disclosure Statement}
AG and SK are employees of Moderna, Inc. and may own stock/stock options in the company. \revise{AMS, YS, GWK, VSB and MHF, EK have ongoing collaborative projects using CUDA-Q which do not alter the scientific integrity of the work presented herein. Other authors declare no conflict of interest.}


\bibliography{references_utf_corrected,achemso-demo}

\end{document}